\documentclass[12pt,a4paper]{article}
\usepackage{jheppub}
\usepackage[T1]{fontenc} 
\usepackage{epsfig} 
\usepackage{amscd}
\usepackage{verbatim}
\usepackage{latexsym}
\usepackage{amsfonts,amsthm,amsmath,mathtools}
\usepackage{amssymb,stmaryrd,textcomp,enumerate,bbm}
\usepackage{color}
\usepackage{xcolor}
\usepackage{booktabs}
\usepackage{float}
\usepackage{graphicx}
\usepackage{slashed}

\notoc 

\title{
$USp(2N_c)$ SQCD$_3$ with antisymmetric:
dualities and symmetry enhancements}

\author[a]{Antonio Amariti,}
\author[b]{Luca Cassia}

\affiliation[a]{INFN Sezione di Milano, Via Celoria 16, Milano, Italy}
\affiliation[b]{Dipartimento di Fisica, Universit\`a di Milano-Bicocca, \\ Piazza della Scienza 3, I-20126 Milano, Italy}
\affiliation[c]{INFN, sezione di Milano-Bicocca, I-20126 Milano, Italy}

\emailAdd{antonio.amariti@mi.infn.it}
\emailAdd{luca.cassia@tuta.io}

\newcommand{\mathe}{\mathrm{e}}
\newcommand{\mathi}{\mathrm{i}}
\newcommand{\mathd}{\mathrm{d}}
\newcommand{\tr}{\mathrm{tr}}
\newcommand{\sgn}{\mathrm{sgn}}
\newcommand{\ind}{\mathrm{ind}}
\newcommand{\vev}{\Phi_\infty}
\newcommand{\cochar}{\boldsymbol{n}}
\newcommand{\fredholm}{\nabla}

\newcommand{\nocontentsline}[3]{}
\newcommand{\tocless}[2]{\bgroup\let\addcontentsline=\nocontentsline#1{#2}\egroup}

\abstract{
We study various aspects of the 4d/3d reduction of  $\mathcal{N}=1$ dualities 
involving $USp(2N_c)$ gauge theories with $2N_f$ fundamentals and one antisymmetric.
We discuss the non-trivial role played by the monopole superpotentials 
in the reduction and obtain new 3d dualities for models with both symplectic 
and unitary gauge groups.
For $N_f=4$ we observe interesting 
webs of dualities and symmetry enhancements, recovering 
and extending some results already appeared in the literature.
}

\begin{document}

\maketitle

\newpage
\tableofcontents
\newpage
\section{Introduction}
%
%
%
%
%
%
%
A fascinating field of research, attracting the interest of both the high energy and the condensed matter 
communities, consists of the 3d analogue of 2d bosonization. This phenomenon can be more generally 
thought of as a limiting case of a broad web of non-supersymmetric 3d dualities 
(see for example \cite{Giombi:2011kc,Aharony:2011jz,Aharony:2012nh,Aharony:2015mjs,Seiberg:2016gmd} for an incomplete list of references). These 3d dualities share many common properties with their 
supersymmetric counterparts, and some attempts to derive them from the supersymmetric case appeared in 
\cite{Jain:2013gza,Gur-Ari:2015pca,Kachru:2016rui,Kachru:2016aon,Aharony:2018pjn}. This provides one of the main motivations for further 
investigations on the supersymmetric side of 3d dualities.

So far most of the non-supersymmetric dualities discussed in the literature refer to gauge theories 
with fundamental matter fields. Recently dualities involving QCD$_3$ with two-index tensor matter 
fields appeared in \cite{Gomis:2017ixy,Cordova:2017vab,Cordova:2017kue}. In the supersymmetric 
case models with two-index tensor matter fields played a relevant role in the generalizations of 4d 
Seiberg duality, starting from the original example of \cite{Kutasov:1995ss}. Furthermore 4d theories 
with fundamental and adjoint matter fields have been recently used as a perturbative description of 
4d $\mathcal{N}=2$ non-lagrangian SCFTs \cite{Maruyoshi:2016aim,Agarwal:2016pjo,Benvenuti:2017kud,Agarwal:2017roi}.

In general 3d dualities involving two-index tensor matter fields have been derived by a circle 
reduction of the 4d cases \cite{Nii:2014jsa,Amariti:2015mva,Amariti:2015vwa} by following the 
prescription of \cite{Aharony:2013dha}. 
This is a stepwise procedure.
The first step consists of finding, on $\mathbb{R}^{1,2} \times S^1$, an effective 3d
description of the 4d duality. 
This can be thought of as a new 3d IR duality, UV completed by the 4d physics.
The final step consists of
taking the 3d limit
and it is accomplished by real mass and Higgs flows. 
By applying this procedure 4d 
Seiberg duality (and its generalizations) reduces to 3d Aharony duality
(and its generalizations).

A different 3d limit was recently considered in \cite{Benini:2017dud}
for the reduction of $USp(2N_c)$ SQCD$_4$.
This led the authors to discover new interesting families of 3d dualities with non-trivial monopole superpotentials
\footnote{See also 
\cite{Dimofte:2012pd,Benvenuti:2016wet,
Amariti:2017gsm,Benvenuti:2017kud,
Amariti:2018gdc}
 for other applications of monopole superpotentials
to 3d $\mathcal{N}=2$ theories.}.
A generalization of these new dualities to the cases with tensorial matter fields has not been obtained yet. 

Here we consider the circle reduction of
$USp(2N_c)$ SQCD$_4$ with one antisymmetric matter field.
The 4d dualities involving this gauge and matter content 
 have been originally discussed in \cite{Intriligator:1995ax}, in the presence of a power law superpotential 
for the antisymmetric field. The generalization of Seiberg duality for such models was obtained and many tests have been performed.
The dynamics of these theories in the absence of the tree level superpotential for the antisymmetric field
has been discussed originally in  \cite{Csaki:1996eu}, in the presence of six fundamentals.
It was shown that this theory confines  in the IR 
without chiral symmetry breaking.

More recently the analysis has been extended to $USp(2N_c)$ gauge theories with
eight fundamentals, an antisymmetric and a set of extra singlets.
Making use of exact mathematical identities for $\mathcal{N}=1$ superconformal indices,
the authors of \cite{Spiridonov:2008zr} constructed a large number of magnetic
duals of this theory, all of which are related by the action of a reflection group
of the type of the $E_7$-root system Weyl group. In \cite{Dimofte:2012pd} it was
then shown that in the rank-1 case these dualities can become a self-duality of
a single theory that exhibits an IR global symmetry enhancement to the full $E_7$
algebra and in \cite{Razamat:2017hda} this result was generalized to arbitrary rank
also showing that the symmetry actually enhances to $E_7\times U(1)$.

In this paper we study the dimensional reduction of these 4d models, finding large classes of new relations and dualities. 
In section \ref{sec:SPas} we study the reduction of $USp(2N_c)$ theories with $2N_f$ fundamentals $Q$, an antisymmetric $A$ and superpotential $W=\tr\,A^{k+1}$.
We generalize the structure of RG flows and 3d dualities already worked out for the 
case without antisymmetric matter.
The main results obtained in this section are highlighted in red in Figure \ref{fig1}.
In section \ref{sec:E7} we study the reduction of the 72 dual phases involving $USp(2N_c)$ gauge groups, 
eight fundamentals and an antisymmetric.
We show that the Weyl group of $E_7$ is still at work, preserving the 4d dualities.
Moreover we show that by real mass flow we can generalize the result of \cite{Dimofte:2012pd}
where the Weyl group of $D_6$ relates two classes of dual theories. On one hand one has $USp(2N_c)$ theories with six fundamentals and one antisymmetric. On the other hand there are $U(N_c)$ gauge theories with four pairs of fundamentals and anti-fundamentals and an adjoint.
By further real mass flow a large web involving $USp/U$ dualities can be constructed. 
We have summarized this construction in Figure \ref{fig2}.
It is interesting to observe that a generalization of the $SU(3)$ global symmetry enhancements, discussed recently in \cite{Gang:2017lsr,Gang:2018wek,Gaiotto:2018yjh,Benini:2018bhk} for $U(1)$ with two fundamentals, is obtained here for $U(N_c)$ with 
two fundamentals and one adjoint.
In section \ref{sec:confining} we study the dimensional reduction of the confining $USp(2 N_c)$ gauge theory with one antisymmetric and six fundamentals. We obtain new 3d confining theories with both $USp(2N_c)$ with fundamentals and one antisymmetric and $U(N_c)$ gauge group
with fundamentals, antifundamentals and one adjoint.
In section \ref{sec:conc} we conclude discussing open questions and future lines of research.
In appendix \ref{sec:app} we review some useful mathematical tools that we used in our analysis.
\\
\\
{\bf Note added :} sections \ref{sec:E7} and \ref{sec:confining} of our work overlap with \cite{Benvenuti:2018bav}, that appeared today.
We are grateful to Sergio Benvenuti for informing us of its results.


%
%
%
%
\section{$2N_f$ fundamentals, one antisymmetric $A$ and $W = \tr \, A^{k+1}$}
\label{sec:SPas}
\begin{figure}
\begin{center}
\begin{tabular}{cc}
\includegraphics[width=0.47\textwidth]{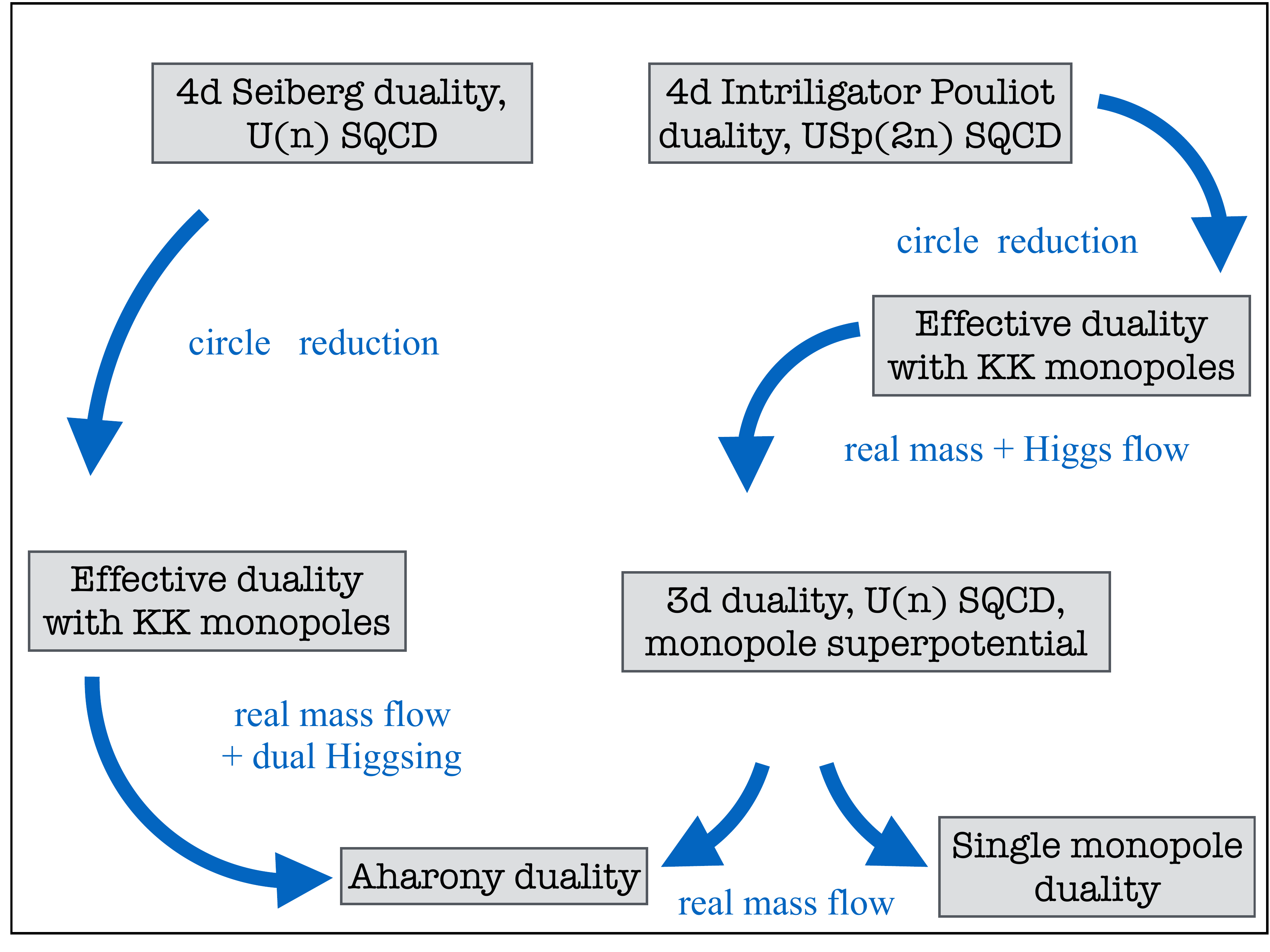}&
\includegraphics[width=0.47\textwidth]{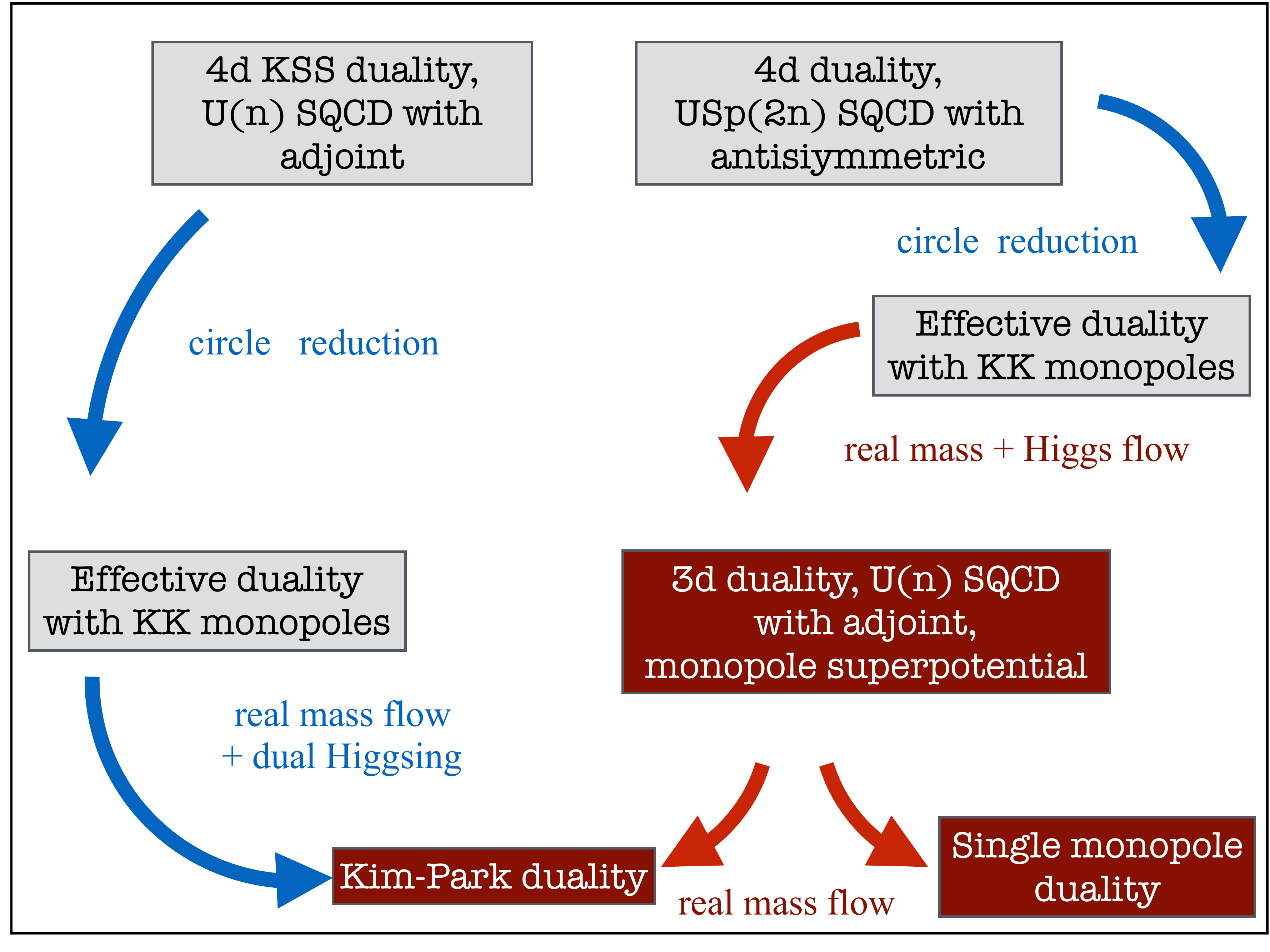}
\\
(a)&(b)
\end{tabular}
\caption{In figure (a) we represented two
different strategies, appeared in the literature, that starting from 4d dualities
led to Aharony duality. 
In the first case one starts from Seiberg duality \cite{Seiberg:1994pq}, reduces on the circle
obtaining a new effective duality \cite{Aharony:2013dha} and then, by a real mass (and a Higgs flow in the dual phase), one recovers Aharony duality \cite{Aharony:1997gp}.
In the second case one starts from the duality of \cite{Intriligator:1995ne}, reduces on a circle and then flows to a new duality \cite{Benini:2017dud}, in the presence of monopole superpotentials.
It is then possible to either flow to the duality with a single term in the monopole superpotential or to the conventional Aharony duality by a real mass flow.
In  figure (b) we discussed the same type of reductions for theories with two-index tensorial matter. The first reduction connects the 4d duality of \cite{Kutasov:1995ss} to the 3d duality of \cite{Kim:2013cma} and it has been studied in \cite{Nii:2014jsa}. In the second case we observe that the 3d duality of \cite{Kim:2013cma} can be recovered also starting from the 4d duality with $USp(2N_c)$ gauge group, with fundamentals and an antisymmetric \cite{Intriligator:1995ax}. This duality has been already reduced on a circle in \cite{Amariti:2015mva,Amariti:2015vwa}, while the other steps have not been performed yet in the literature.  
We highlighted in red these new steps, and they will be the subject of our analysis.}
\label{fig1}
\end{center}
\end{figure}

%
%
%
%
%
In this section we study the reduction to three dimensions of 4d theories with $USp(2N_c)$
gauge group, $2 N_f$ fundamental fields $Q$ and one antisymmetric field $A$, with superpotential
\begin{equation}
\label{eq:W4dAS}
W = \tr \, A^{k+1}.
\end{equation}
This theory has a Seiberg-like dual description \cite{Intriligator:1995ax}, corresponding to
a $USp(2(k(N_f-2)-N_c))$ gauge theory, with $2N_f$ dual fundamentals $q$ and one antisymmetric $a$.
In this case the superpotential of the dual theory is
 \begin{equation}
\label{eq:W4dASdual}
W = \tr \, a^{k+1} + \sum_{j=0}^{k-1} \tr \, M_{k-j-1} q a^j q\,,
\end{equation}
where the generalized mesons $M_j$ are identified with the gauge invariant combinations $Q A^j Q$ and the 
contractions with the symplectic forms are left implicit.
While the reduction on $S^1$ for this model has already been discussed in the literature \cite{Amariti:2015mva,Amariti:2015vwa}, here we will 
study some further flows, constructing dualities involving 3d $U(N_c)$ gauge theories with fundamental flavor, adjoint matter and monopole superpotentials.
These flows are triggered by large real masses and large 
expectation values for the real scalars in the 3d $\mathcal{N}=2$ vector multiplet (a.k.a. Higgs flows).
When $k=1$ they reduce to the flows studied in \cite{Benini:2017dud}
for $USp(2N_c)$ SQCD.
In Figure \ref{fig1} we have depicted the various steps of the dimensional reduction,
 the real mass and Higgs flows studied in the literature for the cases without 
and with matter fields in tensorial representation. We highlighted in red the flows and the models that we study in this section in order to complete the classification.
Observe that here we are not considering dualities 
with CS terms, as the ones considered in 
 \cite{Giveon:2008zn,Benini:2011mf,Niarchos:2008jb,Niarchos:2009aa}.

%
%
%
%
\subsection{The 3d duality with a monopole superpotential}
\label{subsec:monsup}
%
%
%
%
The reduction to three dimensions of this model has already been discussed in the literature in 
\cite{Amariti:2015mva,Amariti:2015vwa}. 
This consists of the stepwise procedure introduced in \cite{Aharony:2013dha} and reviewed in Appendix \ref{APPA}.
Observe that in this case the matter content allows the generation of the  KK superpotential even in the absence of 
superpotential interactions (see Appendix \ref{APPB} for details). This generates an effective duality on $S^1$.
This effective duality on $S^1$ can be deformed into a 3d duality between $USp(2N_c)$ gauge theories as
discussed in \cite{Amariti:2015mva,Amariti:2015vwa}, by a real mass flow.
There is another interesting flow that can be performed, consisting of a mixing between a real mass
and Higgs flow, of the type introduced in \cite{Dimofte:2012pd} and thoroughly investigated in 
\cite{Benvenuti:2016wet,Benini:2017dud}.
Such a flow can be triggered by splitting the real mass $\mu_a$ of the $2N_f$ fundamentals into $N_f$ components $m_a$ and $N_f$ components $n_a$ and considering the following real mass flow 
\begin{equation}
\mu_a \rightarrow m_a + s,
\quad \quad
\mu_{a+N_f} \rightarrow n_a + s,
,\quad \quad
a=1,\dots,N_f
\end{equation}
with $s \rightarrow \infty$. At the same time one needs to consider the
Higgs flow $\sigma_i \rightarrow \sigma_i+s$, with $i=1,\dots,n$.
This leads to a 3d duality between a
$U(N_c)$ theory with  $N_f$ fundamentals $Q$,  $N_f$ anti-fundamentals $\tilde Q$ and one adjoint $X$
with superpotential
\begin{equation}
\label{eq:ASmon}
W = \tr\, X^{k+1} + T_0 + \widetilde T_0 
\end{equation}
and a $U(k(N_f-2)-N_c)$ theory with $N_f$ fundamentals  $q$, $N_f$ anti-fundamentals $\tilde q$ 
and one adjoint $x$, with superpotential
\begin{equation}
\label{eq:ASmond}
W =\tr\,  x^{k+1} + \sum_{j=0}^{k-1} \tr\, M_{k-j-1} q x^j \tilde{q} +  t_0 + \widetilde t_0 
\end{equation}
where the generalized mesons are $M_j = Q X^j \widetilde{Q}$.
Turning on the linear monopole superpotentials in (\ref{eq:ASmon}) and  in (\ref{eq:ASmond}) 
breaks the topological and the axial symmetry and it fixes the R-charges of the fundamentals. 
Furthermore this is consistent with the duality map:
imposing in the electric theory the monopole and the anti-monopole R-charge as
$\Delta_{T_0} = \Delta_{\widetilde {T}_0}=2$ fixes the R-charge of the fundamentals as
\begin{eqnarray}
\label{eq:Rele}
\Delta_Q =\Delta_{\widetilde Q} =  \frac{N_f-2-\Delta_X(N_c-1)}{N_f}
\end{eqnarray}
In the magnetic theory a similar computation gives 
\begin{equation}
\label{eq:Rmag}
\Delta_q = \Delta_{\tilde q} = \frac{2-N_f+\Delta_x(N_c+N_f-1)}{N_f}
\end{equation}
The duality map $\Delta_q = \Delta_X - \Delta_Q$ is satisfied by 
(\ref{eq:Rele}) and (\ref{eq:Rmag}). 

We conclude this discussion with a remark on the linear monopole
superpotentials in (\ref{eq:ASmon}) and in  (\ref{eq:ASmond}) .
The generation of such a superpotential for a $U(N_c)$ gauge theory 
with adjoint matter may appear incorrect, because the adjoint field adds two
further zero modes to the ones carried by the gaugino.
Nevertheless we claimed that the linear superpotential for the bare monopole and 
anti-monopole can be generated. 
This is motivated by the nature of the UV completion of our theory: indeed the $U(N_c)$ 
theory that we are describing so far is UV-completed by a $USp(2N_c)$ 
gauge theory with an antisymmetric field. 
By performing the counting of the zero modes in this setup (see Appendix \ref{app:counting}) one can observe that the antisymmetric field does not lead to any further zero mode in the monopole backgrounds that we are considering, and this signals the presence of the linear monopole superpotentials.
In other words the linear monopoles are inserted in the UV $USp(2N_c)$ theory, where they are perfectly consistent with the zero mode counting, and they modify the Coulomb branch of the IR
$U(N_c)$, treated as an effective theory.

%
%
\subsection{Flowing to Kim-Park duality}
%
%
Starting from the duality above we can flow to the duality of \cite{Kim:2013cma}
as follows.
First we consider a $U(N_c)$ gauge theory with $N_f+2$ flavors and one adjoint with the superpotential 
(\ref{eq:ASmon}).
The dual theory has $U(k N_f-N_c)$ gauge group, $N_f+2$ flavors, one adjoint and the superpotential 
coincides with (\ref{eq:ASmond}).
Then we trigger the flow by shifting the real masses as
\begin{equation}
\label{further}
\begin{array}{ll}
m_{N_f+1} \rightarrow m_{N_f+1} +s,
\quad
&
m_{N_f+2} \rightarrow m_{N_f+2} -s,
\\
n_{N_f+1} \rightarrow n_{N_f+1} -s,
\quad
&
n_{N_f+2} \rightarrow n_{N_f+2} +s,
\end{array}
\end{equation}
with $s\rightarrow \infty$.
In the electric theory we obtain a 3d theory with $U(N_c)$ gauge group, $N_f$ flavors, an adjoint and superpotential $W = \tr\,X^{k+1}$.
In the dual theory the situation is more involved. Indeed in this case some of the components of the original mesons remain massless even if they are not associated to the massless mesons of the electric theory.
These are the $(N_f+1)$-th and the $(N_f+2)$-th diagonal components of 
$M_j$. 
These fields are light gauge singlets in the magnetic theory and their quantum numbers are compatible with the following superpotential interactions
\begin{equation}
W = t_0 \sum_{j=0}^{k-1}  (M_j)_{N_f+1,N_f+1}\, \tr\, x^j 
+
\tilde{t}_0 \sum_{j=0}^{k-1}  (M_j)_{N_f+2,N_f+2}\, \tr\, x^j 
\end{equation}
where $t_0$ and $\tilde t_0$ are the bare monopole and anti-monopole of the $U(k N_f-N_c)$ theory.
The dressed monopoles of the dual theory can be defined as $t_j = t_0\,\tr\,x^j$ and 
$\tilde {t}_j = \tilde t_0\,\tr\,x^j$.
On the other hand the singlets $(M_j)_{N_f+1,N_f+1}$ and $(M_j)_{N_f+2,N_f+2}$ can be identified with 
the dressed monopoles of the electric theory, i.e. $T_j = T_0\,\tr\,X^j$ and $\tilde {T}_j = \tilde T_0\,\tr\,X^j$.
The final form of the superpotential of the dual theory is then
\begin{equation}
W = \tr\, x^{k+1}  + 
\sum_{j=0}^{k-1}  \tr\,  M_{k-j-1} q x^j \tilde{q} 
+
\sum_{j=0}^{k-1} ( t_j T^{k-1-j}  +  \widetilde t_j \widetilde T^{k-1-j})
\end{equation}
reproducing the  dual superpotential of the Kim-Park duality.
%
%
\subsection{Duality with a single monopole superpotential}
\label{subsec:single}
%
%
%
This case can be studied starting with $N_f+1$ flavors and monopole superpotential 
(\ref{eq:ASmon}).
The dual theory in this case has rank $U(k(N_f-1)-N_c)$.
We consider the real mass flow 
\begin{equation}
\label{eq:rmsingle}
m_{N_f+1} = \eta+ s, \quad n_{N_f+1}  = \eta- s
\end{equation}
In the electric theory this real mass flow reduces the number of flavors, and removes the contribution of 
$\widetilde T_0$  to the superpotential (\ref{eq:ASmon}).
This gives a $U(N_c)$ electric theory with $N_f$ flavors and superpotential 
 \begin{equation}
W = \tr \, X^{k+1} + T_0
\end{equation}
Observe that at the level of global currents this superpotential restores a combination of the axial 
$U(1)_A$ and the topological $U(1)_J$ symmetries.

On the dual side the real mass deformation (\ref{eq:rmsingle})   
reduces the number of flavors, while it leaves the rank of the dual group invariant.
The deformation (\ref{eq:rmsingle}) removes the contribution of 
$\widetilde t_0$ to the monopole superpotential and it reduces the number of singlets, from $k (N_f+1)^2$
to $k(N_f^2 + 1)$.
The first $k N_f^2$ components correspond to the generalized mesons of the electric theory, $M_j = Q X^j \widetilde Q$.
The last $k$ components correspond components of the generalized mesons that remain light despite (\ref{eq:rmsingle}).
We refer to these $k$ components
as $S_j$, with $j=0,\dots,k-1$.
Their quantum numbers are compatible with the superpotential
\begin{equation}
\label{eq:magsing}
W = \tr \, x^{k+1} + \sum_{j=0}^{k-1} \tr \,  M_j q x^j \tilde q + t_0 + \sum_{j=0}^{k-1} S_j \widetilde t_j  
\end{equation}
where $\widetilde t_j  = \widetilde t_0\,\tr\,X^j$ are the dressed anti-monopoles of the magnetic theory.
It is then natural to identify $S_j$ with the dressed monopoles of the electric theory.
%
%
%
\subsection{Partition functions }
%
%
%
The sequence of reductions and dualities discussed above can be studied at the level of localization.
The 4d/3d reduction can indeed be analyzed by reducing the identity between the superconformal indices 
relating the 4d duality (see Appendix \ref{APPC} for details).
This provides a relation between the 3d partition functions localized on a squashed three sphere, leading to the identity
\begin{eqnarray}
\label{eq:relASS1}
Z_{USp(2N_c)}(\vec \mu;\tau)
=
\prod_{j=0}^{k-1} 
\prod_{1\leq a<b \leq 2N_f}\Gamma_h(j \tau+\mu_a +\mu_b)
\,
Z_{USp(2\tilde N_c)}(\vec{\widetilde \mu};\tau)
\end{eqnarray}
where the parameters of the electric theory are related by the constraint (referred in \ref{APPC}
as balancing condition)
\begin{equation}
\label{eq:BCASS1}
\sum_{a=1}^{2N_f} \mu_a = 2 (\omega(N_f-2)-(N_c-1) \tau)
\end{equation}
The parameter $\tau$ is constrained here by the superpotential, it corresponds to $\tau  = \frac{\omega}{k+1}$.
Indeed the antisymmetric field in this case is not charged under any non-R global symmetry because of the superpotential.
The parameter $\tau$ is purely imaginary and corresponds to the 
$R$-charge of the antisymmetric field.
The rank of the dual group is $\widetilde N_c = k(N_f-2)-N_c$.
The parameters $\tilde \mu_a$ are related to the electric ones by the duality map
$\widetilde \mu_a = \tau - \mu_a$.

Starting from the relation (\ref{eq:relASS1}) we can shift the scalars $\sigma_i \rightarrow \sigma_i+s$ and
consider the real mass flow $\mu_a = m_a+  s,\,\mu_{a+N_f} = n_{a} - s$ 
for $1\leq a \leq N_f$.
This does not affect the balancing condition (\ref{eq:BCASS1}) but it leads to the relation
\begin{eqnarray}
\label{eq:monsup}
Z_{U(N_c)}(\vec m;\vec n;\tau)=
\prod_{j=0}^{k-1} 
\prod_{a,b=1}^{N_f}\Gamma_h(j \tau+m_a +n_b)
\,
Z_{U(\widetilde N_c)}(\vec {\widetilde m};\vec {\widetilde n};\tau)
\end{eqnarray}
where $\widetilde m_a = \tau - m_a$,  $\widetilde n_a = \tau - n_a$.
Relation (\ref{eq:monsup}) represents the equivalence between the partition functions of the models
discussed in subsection \ref{subsec:monsup}. The presence of the monopole superpotential is 
encoded in the constraint (\ref{eq:BCASS1}).
Observe that the presence of this constraint breaks the otherwise non-anomalous axial symmetry.
The breaking of $U(1)_J$ is encoded in the absence of an FI.

We can also reproduce the flow
 to the Kim-Park duality on the partition function.
The final relation was originally proved in \cite{Amariti:2014iza} by considering the reduction of KSS duality on $S^1$.
The flow from the effective duality on $S^1$ to the Kim-Park duality required a Higgsing in the dual phase.
This Higgsing led to a product of gauge groups. One of them represented the dual gauge group while the other needed
to be dualized to  set of singlet, corresponding to the electric monopoles acting as singlets in the dual phase.
On the partition function this dualization was possible because of an exact relation at the level of the 3d partition function, discussed in \cite{VdB}. Interestingly this relation played a relevant role recently in \cite{Aghaei:2017xqe}, in the reduction of AD theories to 3d.

Here we arrive at the same final identity proven in \cite{Amariti:2014iza} by following a different strategy.
First we consider the monopole duality with $N_f+2$ flavors. Then, on the electric side we shift the flavors as 
discussed in  (\ref{further}).
The final identity is 
\begin{eqnarray}
\label{eq:KimPArk}
Z_{U(N_c)}(\vec m;\vec n;\tau;\Lambda)
&=&
\prod_{j=0}^{k-1} 
\Gamma_h \Big(\pm \frac{\Lambda}{2} + \omega N_f+ (j-N_c+1) \tau - \sum_{a=1}^{N_f} \frac{m_a+n_a}{2} \Big)
\nonumber \\
&\times &
\prod_{a,b=1}^{N_f}\Gamma_h(j \tau+m_a +n_b)
Z_{U(\widetilde N_c)}(\vec {\widetilde m};\vec {\widetilde n};\tau;-\Lambda)
\end{eqnarray}
Observe that the balancing condition becomes
\begin{equation}
\label{eq:BCASS1bis}
\sum_{a=1}^{N_f} (m_a+n_a) + 2(m_{N_f+1}+m_{N_f+2}) = 2 (\omega N_f-(N_c-1) \tau)
\end{equation}
The parameters $m_{N_f+1}$ and $m_{N_f+2}$ are free, 
signaling the absence of a balancing condition on the 
mass parameters $m_a$ and $n_a$. On the physical side 
the combinations $m_{N_f+1}+m_{N_f+2}$ represents the presence of an axial symmetry 
while the parameter
$\Lambda = 2(m_{N_f+1}-m_{N_f+2})$ represents the Fayet--Iliopoulos (FI) term. 
It indeed appears as $e^{i \pi \Lambda \sum \sigma_i}$
on the LHS of (\ref{eq:KimPArk}) and with an opposite side on the RHS.
The first term on the RHS of (\ref{eq:KimPArk}) represents the contribution of the dressed monopoles
of the electric theory acting as singlets on the dual side.

We conclude the analysis by considering the 
real mass flow studied in sub-section \ref{subsec:single}, leading to the theory with a single monopole superpotential.
In this case we consider the electric theory with $N_f+1$ flavors and deform it as in (\ref{eq:rmsingle}).
The balancing condition becomes
\begin{equation}
\label{eq:bcsingle}
2\tau(N_c-1) + \sum_{a=1}^{N_f} (m_a+n_a) +2 \eta= 2 \omega (N_f-1)
\end{equation}
The presence of $\eta$ in (\ref{eq:bcsingle}) signals the fact that an extra abelian symmetry has been generated by the real mass flow.
To understand the origin of this symmetry one can look at the final identity relating the partition functions of the electric and of the magnetic theory.
In this case case we have
\begin{eqnarray}
\label{eq:monsing}
&&
Z_{U(N_c)}(\vec m;\vec n;\tau;\omega-\eta) 
=
\mathe^{\frac{\pi\mathi}{2} k \sum _{a=1}^{N_f} \left(m_a^2-n_a^2\right)}
\prod_{a,b=1}^{N_f}\Gamma_h(j\tau+m_a +n_b)
\\
\times
&&
\prod_{j=0}^{k-1} \Gamma_h \big(\eta+ \omega (N_f-1)+\tau(j-N_c+1)
\! -
\!
\frac{1}{2}\sum_{a=1}^{N_f} (m_a+n_a)
\big)
Z_{U(\tilde N_c)}(\vec {\widetilde m};\vec {\widetilde n};\tau;\tau \!-\!\omega\!-\!\eta)
\nonumber 
\end{eqnarray}
where $\widetilde m_a = \tau- m_a$, $\widetilde n_a = \tau- n_a$ and  $\widetilde N_c = k(N_f-1)-N_c$.
The result matches the expectations from the field theory analysis. Indeed one can read the 
 charges of the singlets from the partition function and check that they coincide with the ones obtained for the singlets $S_{j}$ in the superpotential (\ref{eq:magsing})
%
%
%
%
%
%
%
\section{Eight fundamentals and $E_7$ symmetry}
\label{sec:E7}
%
%
%
%
%
%
%
In this section we re-consider the gauge and field content discussed above, i.e.,
supersymmetric gauge theories with a symplectic gauge group, fundamentals $Q$
(here we restrict to $2N_f = 8$) and one anti-symmetric tensor matter field $A$. 
However, here we have models without a power law superpotential for the field $A$.
These theories have been analyzed in \cite{Spiridonov:2008zr,Dimofte:2012pd,Razamat:2017hda}
where it has been shown that these models present an IR enhancement of the
global symmetry to $E_7 \times U(1)$. In the following we will study the reduction
of these models to 3d, showing the appearance of monopole superpotentials and
constructing an intricate web of dualities. These new dualities generalize the 3d
$SU(2)/U(1)$ duality discovered in \cite{Dimofte:2012pd} to a $USp(2N_c)/U(N_c)$
duality. We also emphasize the key role played by monopole superpotentials in the
analysis and check many of the claims by testing them with the three sphere
partition function.
%
\subsection{The 4d theory}
\label{subsec:4dE7}
%
%
%
%
The 4d theories have been largely discussed in 
{\cite{Spiridonov:2008zr,Razamat:2017hda}
and here we will just briefly review some of the main aspects
of these models necessary for our analysis.
One can divide the 4d theories into 4 classes, depending on the global symmetry and on the presence of singlets
that can be added without further breaking of the flavor symmetry.
The first two classes, ({\bf A}) and ({\bf B}) have a classically unbroken $SU(8)$ global symmetry, while the other two classes, ({\bf C}) and ({\bf D}) have
a smaller $SU(4) \times SU(4)$  classical global symmetry. In the following we discuss some of the salient features of these theories.
\begin{itemize}
\item [({\bf A})]  
The theory has a global $SU(8) \times U(1) \times U(1)_R$ symmetry group and 
the fields transform under the gauge and global symmetries as 
\begin{equation}
\begin{array}{c|cccc}
       & USp(2N_c)  &   SU(8)  &  U(1)' &    U(1)_R  \\
       \hline
Q    &   2N_c          &      8      &  \frac{1-N_c}{4}&  R_Q\\
A   & N_c(2N_c-1)-1  &      1      &    1& R_A
\end{array}
\end{equation}
The anomaly freedom of the $R$-symmetry 
imposes the constraint 
\begin{equation}
(N_c+1) + (N_c-1) (R_A-1) + 4(R_Q-1) = 0  
\end{equation}
\item [({\bf B})]
 This is a Seiberg dual phase, the global symmetry group visible in the lagrangian is maximal, i.e. 
$SU(8) \times U(1) \times U(1)_R$. This dual theory has $N_c$ mesons in the two index antisymmetric 
representation of the non-abelian $SU(8)$. These 28 dimensional mesons act as singlets in the dual 
phase and they can be expressed in terms of the matter fields of the electric phase as
\begin{equation}
M_{rs}^{(j)} = Q_r Q_s A^{N_c-1-j}
\quad 
\text{with}
\quad 
1\leq r <s \leq 8
\quad 
\text{and}
\quad  j=0,\dots,N_c-1
\end{equation} 
The eight dual fundamentals $q$ and the dual antisymmetric $a$ interact with the meson
through the superpotential 
\begin{equation}
\label{Wmag}
W_{B} = \sum_{j=0}^{N_c-1} \tr\, M^{(j)} q q a^{j} 
\end{equation}
\item [({\bf C})]
In this case the global $SU(8)$ symmetry is explicitly broken to $SU(4) \times SU(4) \times U(1)_B$,
where the subscript $B$ in the abelian factor indicates
that this symmetry acts like a baryonic symmetry giving an opposite charge to the fundamentals
of the two $SU(4)$ factors.
There are $N_c$ mesons  $M^{(j)}$ in the $4 \times \overline 4$ representation of the non-abelian symmetry group and 
there is a  superpotential  
\begin{equation}
W_{C} = \sum_{j=0}^{N_c-1} \tr\, M^{(j)} q p a^{j}
\end{equation}
where the four anti-fundamentals $q$ refer to the first $SU(4)$ factor and the four fundamentals
$p$ refer to the second $SU(4)$ factor.
Up to permutations there are $\frac12\binom{8}{4}=35$ inequivalent theories,
having the same field content in terms of gauge group and 
charged matter. All these models are claimed to be dual to the ones presented in ({\bf A}) and ({\bf B}).
\item [({\bf D})]
There is a second family of theories with an $SU(4)^2$ manifest global symmetry group. This theory 
has two types of mesons, 
each one in the antisymmetric representation of one of the two $SU(4)$ factors.
Referring to these mesons as $M^{(j)}$ and $N^{(j)}$
the superpotential becomes
\begin{equation}
\label{WII}
W_D = \sum_{j=0}^{N_c-1}  \tr\,\big( M^{(j)} q q a^{j} + N^{(j)} p p a^{j} \big)
\end{equation}
Also in this case there are 35 inequivalent theories and they are claimed to be dual to the ones discussed in ({\bf A}), ({\bf B}) and ({\bf C}).
\end{itemize}
The duality among these 72 models has been claimed in \cite{Spiridonov:2008zr}, by
use of the integral identities of \cite{Rains} between their superconformal indices.
These identities correspond to the action of the Weyl group of $E_7$ on the chemical
potentials associated to the global symmetries.
One can then imagine that the set of 72 dual theories forms an orbit for the action
of the Weyl group of $E_7$ with stabilizer the parabolic subgroup $S_8\cong W(A_7)$
corresponding to the manifest global symmetry of the lagrangian which acts by
permutation of the fundamental fields. The size of the orbit is then given by the
ratio of the orders of the two groups, which is precisely the number 72.
In the case of even $N_c$ it has been also observed \cite{Razamat:2017hda} that all the models can be deformed in such a way that one deals with 72 self dual phases.
In such cases the (self)-duality group enhances the $SU(8) \times U(1)$ global symmetry to $E_7 \times U(1)$.
%
%
%
%
\subsection{Reduction to 3d}
%
%
%
%

The models described above can be reduced to 3d by a circle compactification. 
The prescription of \cite{Aharony:2013dha},
 reviewed in appendix \ref{APPA}, is necessary in order to preserve the
duality among the different 72 phases.
We reduce the spectrum and the interactions of each phase and then add the KK monopole superpotential
(see \ref{APPB} for details).
The presence of the KK monopole superpotential imposes further constraints on the 3d real masses 
of the matter fields.
In this case the constraint is
\begin{equation}
\label{eq:etaconstr}
(N_c+1) + (N_c-1) (\Delta_A-1) + 4 (\Delta_Q -1) = 0  
\end{equation}
where $\Delta_Q$ and $\Delta_A$ are the 3d R-charges of the fundamentals and 
of the antisymmetric respectively.
Observe that the constraint (\ref{eq:etaconstr})
is equivalent to the one imposed in 4d by the
anomaly freedom of the R-current.
The KK monopole superpotential constrains the global symmetries as well, 
preventing the generation of possible axial symmetries.

This procedure preserves the duality among the 72 $USp(2 N_c)$ theories 
with eight fundamentals and an antisymmetric. This can indeed be thought of
as a duality between 3d effective theories.
This claim can be tested by reducing the identities between the
superconformal indices to identities between the three sphere partition functions. 
The final identities already appeared in the literature in \cite{VdB}
(see  \cite{Rains:2006dfy} for a seminal work).
Here we translate in a physical language many of the 
results of  \cite{VdB}, deriving an interesting set of new 3d $\mathcal{N}=2$ dualities.
The starting point consists of considering the exact mathematical identity 
\begin{equation}
\label{E7}
Z_{USp(2N_c)}(\vec \mu;\tau)
=
\prod_{j=0}^{N_c-1} 
 \prod_{1 \leq r < s \leq 4} \Gamma_h (j \tau + \mu_r + \mu_s)
 \prod_{5 \leq r < s \leq 8} \Gamma_h (j \tau + \mu_r + \mu_s)
Z_{USp(2N_c)}(\vec  {\widetilde \mu};\tau)
\end{equation}
where
we defined  
$\widetilde{\mu }= \{\mu_r+\zeta, \mu_{r+4}-\zeta\}$ for $r=1,\dots,4$
and
\begin{equation}
\label{eq:zeta1}
2 \zeta = \sum_{r=5}^8 \mu_r - 2 \omega +(N_c-1) \tau =
-\sum_{r=1}^4 \mu_r + 2 \omega - (N_c-1) \tau  
\end{equation}
This identity holds provided the constraint
\begin{equation}
\label{balancing}
2 (N_c-1) \tau + \sum_{r=1}^{8} \mu_r = 4 \omega
\end{equation}
is  imposed on the mass parameters $\mu_r$ and $\tau$.
The  relation (\ref{E7}) together with the invariance of the integral 
under permutations of the eight $\mu_r$ variables provides the 
invariance under the action of $W(E_7)$ \cite{Rains:2006dfy,VdB}.
Observe that (\ref{E7}) can be viewed as a master relation and that all possible other dualities 
can be proved by alternating (\ref{E7}) and permutation. 
 
For example the duality between the two models with a  manifest $SU(8)$ global symmetry follows from (\ref{E7}). 
It is obtained by alternating the  transformation (\ref{E7}) and three permutations.
More precisely one first applies  (\ref{E7}) to the $\mu_i$ ordered as above.
Then one permutes the $\tilde \mu$ variables exchanging $\widetilde \mu_{3}$
and $\widetilde \mu_{4}$ with $\widetilde \mu_{5}$
and $\widetilde \mu_{6}$ and apply  (\ref{E7})  again. The last permutation corresponds 
to exchange $\widetilde {\widetilde \mu}_{3}$ and $\widetilde {\widetilde \mu}_{4}$
with $\widetilde {\widetilde \mu}_{7}$ and $\widetilde {\widetilde \mu}_{8}$
and apply (\ref{E7})  for the third time. 
Observe that each time we apply the transformation (\ref{E7}) we generate $12 N_c$ new mesons,
corresponding to $N_c$ times the two antisymmetric representation of each $SU(4)$ global symmetry group.
However the duality with the manifest $SU(8)$ global symmetry has $N_c$ mesons in the antisymmetric
representation of $SU(8)$ corresponding to $28 N_c$ components.
One can observe explicitly that the extra $8 N_c$ components are pairwise massive and eliminate them
on the integral identity by iterating the relation (\ref{holom}).
The final relation that one obtains is
\begin{equation}
\label{maximal}
Z_{USp(2N_c)}(\vec \mu;\tau) = 
\prod_{j=0}^{N_c-1} \prod_{r<s} \Gamma_{h}(j \tau + \mu_r +\mu_s)
\,
Z_{USp(2N_c)}(\vec {\widetilde \mu};\tau) 
\end{equation}
with $\widetilde  \mu_r = \omega-\frac{N_c-1}{2} \tau -\mu_r$

It is interesting to observe that this last duality reduces to 
one of the cases discussed in Section \ref{sec:SPas}
if we add to the superpotential 
the deformation $W = A^{N_c+1}$. This superpotential deformation
corresponds on the dual side to the contribution 
$W = a^{N_c+1}$ for the dual antisymmetric field.
This deformation breaks the $U(1)'$ symmetry and it forces $\tau = \frac{2\omega}{N_c+1}$.
It corresponds to the reduction of the duality of \cite{Intriligator:1995ff} studied in \cite{Amariti:2015vwa}.
Indeed in this case the dual mode must have $USp(2(k(N_f-2)-N_c))$ gauge symmetry, where here $k=N_c$ and
$N_f=4$. The identity (\ref{maximal}) corresponds to the one obtained in \cite{Amariti:2015vwa}
if the actual value of $\tau$ is inserted in the balancing condition.
\\

We conclude by observing that the global symmetry 
of the integrals can enhance to $W(E_7)$ if $N_c$ is even, similarly to the 
4d case.
First one adds the superpotential 
\begin{equation}
\label{additional}
\Delta W = \sum_{j=0}^{\frac{N_c}{2}-1} \tr\, M^{(j)} Q Q A^{j} + \sum_{j=2}^{N_c} \beta_j \tr\, A^{j}
\end{equation}
Then one observes that all the theories are self dual if this deformation is provided.
At the integral level this signals the fact the we must have an integral identity 
in which each phase is just a re-parametrization of the real masses, without further
uncharged matter fields distinguishing the different phases. This corresponds to an enhanced symmetry
and not to a duality.
This can be proven on the partition function by showing that the generator of the Weyl reflection
that does not correspond to a permutation is just a re-parametrization of the masses.
The further generator is the one generating the identity (\ref{E7}). The addition of the superpotential (\ref{additional})
corresponds to multiplying the identity (\ref{E7}) by the terms
\begin{equation}
\prod_{r<s} \prod_{j=0}^{\frac{N_c}{2}-1}
\Gamma_h(2 \omega - j \tau - \mu_r -\mu_s) \times \prod_{j=2}^{N_c} \Gamma_h(2 \omega-j \tau)
\end{equation}
On the LHS some of the terms simplify against the contributions of the mesons in (\ref{E7}).
The mesonic contributions that do not simplify correspond to the term 
\begin{equation}
\prod_{
\tiny
\begin{array}{c}
r=1,\dots,4 \\
s=5,\dots,8 \\
j=0,\dots,\frac{N_c}{2}-1
\end{array}}
\!\!\!\!\!\!\!\!\!\!
\Gamma_h(2 \omega - j \tau - \mu_r -\mu_s)
\!\!\!
\prod_{j=\frac{N_c}{2}}^{N_c-1} 
\big( \!\!\!\!\!
\prod_{1\leq r<s \leq 4}  
\!\!\!\!\!
\Gamma_h(2 \omega - j \tau - \mu_r -\mu_s) 
 \!\!\!\!\! 
\prod_{5\leq r<s \leq 8} \Gamma_h(2 \omega - j \tau - \mu_r -\mu_s)
\big)
\end{equation}
We can substitute in this expression the real masses $\widetilde{\mu}_r= \mu_r -\zeta$ for $r=1,\dots,4$
and $\widetilde{\mu}_r = \mu_r +\zeta$ for $r=5,\dots,8$ and when necessary plug in the condition
(\ref{eq:zeta1}).
The final result is that in the dual theory we remain with the same pre-factor added to the LHS of (\ref{E7}) in terms of the $\widetilde{\mu}$
masses.
\begin{equation}
\prod_{r<s} \prod_{j=0}^{\frac{N_c}{2}-1}
\Gamma_h(2 \omega - j \tau - \widetilde{\mu}_r -\widetilde{\mu}_s) \times \prod_{j=2}^{N_c} \Gamma_h(2 \omega-j \tau)
\end{equation}
Recalling that the identity (\ref{E7}) together with the permutations of $\mu$ generates the group $W(E_7)$, proves that this is a symmetry of the integrals and 
supports the claim that the model with the superpotential (\ref{additional}) enhances the global symmetry group to $E_7$.
%
%
%
\subsection{Real mass flows: $USp(2N_c)$ models and the action of $D_6$}
%
%
%
%
The next step consists of removing the KK monopole superpotential
to obtain conventional 3d $\mathcal{N}=2$ models.
This is done by integrating out some flavors, i.e. by assigning
a large real mass to them.
When this procedure is done consistently on a pair of 
dual theories such a duality can be preserved in the 3d limit \cite{Aharony:2013dha}.
In the case discussed here we have a set of dual phases connected by a 
larger symmetry group than the one expected from the action.
The concepts of global symmetry and of duality are here strongly 
connected, and depending on the details of the model the action 
of a duality group can become the action of a global symmetry group.
The group underlining this web of theories is the Weyl group of 
$E_7$.

We have just reviewed its action on  the complex combinations
$\mu$, representing the real masses and the R-charges of the matter fields.
By a real mass flow the $W(E_7)$ symmetry group is generically broken to a subgroup.
This subgroup is associated to the action of the global symmetry 
group of the IR theory.
In the following we study an explicit realization of such a mechanism 
by assigning an opposite large mass to two fundamentals in the 
$USp(2N_c)$ theory with one antisymmetric and  eight fundamentals.
This assignment must be done consistently with the duality map (\ref{eq:zeta1}).
For example we can assign the large masses as $\mu_7=M+\xi_7$ and $\mu_8=-M+\xi_8$,
with $M>0$. The parameters $\xi_{7,8}$ can be eliminated after using the original constraints
between the real masses (\ref{balancing}), signaling the fact that we will 
be left with a set of unconstrained real masses at the end of the flow. This 
is consistent with the generation of an extra symmetry, constrained before by the presence 
of the KK monopole superpotential. This is similar to the generation of the axial symmetry in the 
reduction of 4d Seiberg duality to 3d Aharony duality.
This signals the fact that the monopole superpotential vanishes as well.

Let us consider the effect of such a mass deformation in 
one of the dual phases introduced above. 
We proceed as follows: we pick up a pair of dual models, 
treating them as a representative of the duality.
We discuss the real mass flow for this pair of dual models and then
extract the necessary information to reconstruct the full duality symmetry.
We first study the following reduction between a pair of dual models:
\begin{itemize}
\item On the electric side we consider 
3d $\mathcal{N}=2$ $USp(2N_c)$ gauge theory with an anti-symmetric and eight
fundamental quarks.
In this case the electric theory, after the real mass flow, becomes $USp(2N_c)$ with 
six fundamentals and one antisymmetric.
The real mass parameters are unconstrained.
\item On the magnetic side we consider 3d $\mathcal{N}=2$ $USp(2N_c)$ gauge theory with an anti-symmetric, eight fundamental quarks and superpotential $W_{D}$ in (\ref{WII}).
The situation in this dual theory is more interesting. The mesons $M^{(j)}$ are light and survive in the low energy spectrum.
The other mesons that survive are the components $N_{56}^{(j)}$ and $N_{78}^{(j)}$.
While the first set corresponds to the $N_c$ generalized mesons of the $SU(2) \subset SU(4)$ original flavor symmetry, the second set is
associated to $N_c$ new singlets, that we denote as $T_j$.  By looking at the charges of these singlets,
they correspond to the dressed electric monopoles, combinations of the bare monopole $T_0$ of the
electric theory with powers of the antisymmetric field $X$, i.e., $T_j = T_0 \tr\, A^j$.
This is consistent with a superpotential of the form 
\begin{equation}
W_{T_j} = \sum_{j=1}^{N_c} T_{N_c-j} \tr\, p_5 p_6 a^{j-1}  
\end{equation}
The interaction with the other meson  $N_{56}^{(j)}$ vanishes, because the fields $p_7$ and $p_8$ are massive in the dual phase.
Nevertheless the charges of this meson are consistent 
\footnote{Observe that further checks are necessary to prove the existence of such an interaction.}
with the interaction
\begin{equation}
W_{N_{56}} = \sum_{j=1}^{N_c} N_{56}^{(j)} t_{j-1} 
\end{equation}
where $t_j$ represents the dressed monopole of the magnetic theory $t_j = t_0 \tr\, a^j$.
The final structure of the superpotential of the dual theory is
\begin{equation}
W = \sum_{j=1}^{N_c} 
\bigg(
 \tr\, M^{(j)} q q a^{j-1} 
+
T_{N_c-j}\, \tr\, p_5 p_6 a^{j-1} 
+
N_{56}^{(j)} t_{j-1} 
\bigg)
\end{equation}
\end{itemize}

\noindent One can also follow this real mass flow on the partition function.
The duality is preserved if the divergent terms coincide in the relation (\ref{E7})
after the infinite shifts are performed.
The real mass flow is performed by using the relation (\ref{iout}).
The divergent term in the electric partition function is a phase 
$e^{\frac{i \pi}{\omega_1 \omega_2} \phi_e}$
with
\begin{equation}
\phi_e
=
2 N_c \left(2 M+\xi _7-\xi _8\right) \left(\xi _7+\xi _8-2 \omega \right)
\end{equation}
In the dual model there are two phases contributing to the divergent term. The first comes
from the mesons and the second one from the dual fundamentals, $\phi_m = \phi_{mes.} + \phi_{fund.}$.
They are
\begin{eqnarray}
\phi_{mes.} &=& \,\,\,\,\,2 N_c \left(2 M+\xi _7-\xi _8\right) \left(\mu _5+\mu _6+(N_c-1) \tau +\xi _7+\xi _8 -2 \omega \right)
\nonumber \\
\phi_{fund.} &=&-2 N_c \left(2 M+\xi _7-\xi _8\right) \left(\mu _5+\mu _6+(N_c-1) \tau \right)
\end{eqnarray}
One can check that $\phi_e = \phi_m$ leading to the equality
\begin{eqnarray}
\label{D6a}
Z_{USp(2N_c)_0}(\vec \mu;\tau)
&=&
\prod_{j=0}^{N_c-1} 
 \prod_{1 \leq r < s \leq 4} \Gamma_h (j \tau + \mu_r + \mu_s)
 \Gamma_h (j \tau + \mu_5 + \mu_6) 
 \\
 &\times& 
\Gamma_h\left(4\omega  -(2N_c-2+j)\tau -\sum_{r=1}^{6} \mu_r \right)
\,
Z_{USp(2N_c)_0}(\vec{\widetilde \mu};\tau)
 \nonumber
\end{eqnarray}
where $\tilde \mu_r = \mu_r + \zeta$ for $r=1,\dots,4$ and  
$\tilde \mu_r = \mu_r - \zeta$ for $r=5,6$ and
$2\zeta = 2 \omega -\sum_{r=1}^{4} \mu_r -(N_c-1) \tau$.
Observe that this relation corresponds to {\bf Theorem 5.6.11} of \cite{VdB}
after applying the identity $\Gamma_h(2\omega -x) \Gamma_h(x)=1$ on the last term in 
(\ref{D6a}).
As explained there, the integrals have an $W(D_6)$ symmetry, generated by the combined action of the 
permutation of the parameters $\mu_r$ and by the transformation (\ref{D6a}).
\\
\\
One can also study the real mass flow triggered by
$\mu_7=M+\xi_7$ and $\mu_8=-M+\xi_8$ on the dual model that preserves the 
whole $SU(8)$ flavor symmetry, identified by the superpotential $W_{B}$ in (\ref{Wmag}).
In this case some of the components of the mesons of the dual theory are massive, while there are 
$N_c$ singlets,
associated to the combination $Q_7 Q_8 A^{j}$, massless in the dual theory, that do not give rise to  
any generalized meson in the dual theory. They correspond to the dressed monopole operators of the electric theory, $T_0\, \tr\, A^j$,
acting as singlet in the dual phase. By looking at the charges of these singlets under the global symmetry one can observe that
there is a superpotential interaction compatible with the presence of these fields. The interaction is
\begin{equation}
\label{wmono}
W = \sum_{j=0}^{N_c-1} T_{j} \, t_{N_c-1-j}
\end{equation}
where $t_j = t_0\, \tr\, a^j$ are the dressed monopoles of the dual theory.
We can perform some checks of this duality.
\begin{itemize}
\item
As a first consistency check we observe that if the antisymmetric acquires a mass term and it is integrated out, the superpotential
(\ref{wmono}) corresponds to the one expected for the Aharony duality for $USp(2N_c)$ with six fundamentals and $N_c=1$.
\item
Another check of the duality just stated consists of studying the real mass flow on the partition function.
Proceeding as above we arrive at the relation:
\begin{eqnarray}
\label{identityD6max}
Z_{USp(2N_c)}(\vec \mu;\tau)
&=&
Z_{USp(2N_c)}(\vec { \tilde \mu};\tau)
\prod_{j=0}^{N_c-1} 
\prod_{1\leq r<s \leq 6} \Gamma_{h}(j \tau + \mu_r +\mu_s)
\nonumber \\
&\times &
\Gamma_{h}\Big(4 \omega - (2N_c-2+j) \tau -\sum_{r=1}^{6} \mu_r\Big)
\end{eqnarray}
with $\tilde \mu_r=\omega-\frac{n-1}{2} \tau - \mu_r$, for $r=1,\dots,6$.
From this relation we can read the real mass $\hat{m}_{ele}^{(j)}$ of the $j$-th electric
dressed monopole  $\hat{m}_{ele}^{(j)}=4 \omega - (2N_c-2+j) \tau -\sum_{r=1}^{6} \mu_r$.
The $j$-th magnetic dressed monopole has real mass
$\hat{m}_{mag}^{(j)}=4 \omega - (2N_c-2+j) \tau -\sum_{r=1}^{6} \tilde {\mu}_r$.
It follows that $\hat{m}_{ele}^{(j)} +\hat{m}_{mag}^{(N_c-1-j)} = 2 \omega$, corresponding to the
constraint imposed by the superpotential (\ref{wmono}).
\item
As a last check we can show that also in this case the duality reduces to the one studied in \cite{Amariti:2015vwa} if we add a superpotential 
$W = \tr\, A^{k+1}$ to the antisymmetric field in the electric side and an analogous one on the dual side.
In this case the dual theory is expected to have $USp(2(k(N_f-1)-N_c))$  gauge symmetry, where 
$N_f=3$ and $k=N_c$.  
The identity (\ref{identityD6max}) coincides with the one derived in \cite{Amariti:2015vwa}
after the actual value of $\tau$ is inserted in the balancing condition. 
\end{itemize}
~
\\
We can modify this duality to
a self duality if $N_c $ is even. On the field theory side this can be done by flipping 
half of the singlets $T_{j}$ and $M^{(j)}$. On the partition function
this is done equivalently by multiplying both sides of the identity (\ref{D6a}) by 
\begin{equation}
 \prod_{j=0}^{\frac{N_c}{2}-1} 
 \frac{1}{
  \prod_{1 \leq r < s \leq 6} 
  \Gamma_h (j \tau + \mu_r + \mu_s)
\Gamma_{h}\Big(4 \omega - (2N_c-2+j) \tau -\sum_{r=1}^{6} \mu_r\Big)
}
\end{equation}
and then using $\Gamma_{h}(x) \Gamma_h(2\omega-x)=1$ together 
with the balancing condition.
By proceeding in a similar fashion one can work out the explicit matter content and
superpotentials of the other possible phases.
We can also count the number of dual phases: there are $|W(D_6)|/|W(A_5)|=32$ dual phases.
This corresponds to the calculation performed in \cite{Dimofte:2012pd}.
We will further comment on this number in sub-section \ref{subsec:general-scheme}, 
where we will explain its
algebraic origin and study further real mass flows, constructing a full duality web.
%
%
%
%
%
%
\subsection{Higgs flow and new $U(N_c)$/$USp(2N_c)$ dualities}
%
%
%
%
%
%

The action of the Weyl group of $D_6$ on the real mass parameters can be also obtained by engineering a different flow on the original duality.
This is essentially the same type of flow studied in \cite{Dimofte:2012pd}, that led the authors to conjecture an $SU(2)/U(1)$ duality.
Similarly here we will claim the existence of an $USp(2N_c)/U(N_c)$ duality.
In this case the real mass flow has to be supplemented by an Higgs flow, 
that has indeed the effect of
breaking the $USp(2N_c)$ gauge symmetry to $U(N_c)$.
In order to study such a flow here we first order the masses $\mu_i$ as $(m_1,m_2,m_3,n_4,n_1,n_2,n_3,m_4)$
and then we consider the infinite shifts $m_r+M$ and $n_r-M$ for $r=1,\dots,4$ and $M>0$.
The change in the labels of the masses is done only to match with the notations of \cite{VdB}.

The Higgs flow is triggered by assigning a vev to the scalar $\sigma$ 
in the vector multiplet. This is equivalent to consider the shift $\sigma_i \rightarrow \sigma_i + M$.
This shift breaks $USp(2N_c)$ with eight fundamentals and an antisymmetric 
into 
$U(N_c)$ with four flavors and one adjoint.
The mass parameters are still constrained by the relation
\begin{equation}
\label{b2}
2(N_c-1) \tau + \sum_{r=1}^{4} (m_r +n_r) = 4 \omega
\end{equation}
signaling the presence of a superpotential  
\begin{equation}
\label{monW}
W = T_0+ \widetilde{T}_0
\end{equation}
This is the type of flow from symplectic to unitary gauge groups
studied in \cite{Dimofte:2012pd,Benini:2017dud}, and indeed the superpotential (\ref{monW}) is the generalization 
of the one obtained in the case without antisymmetric matter.
As already observed in sub-section \ref{subsec:monsup} 
this superpotential is generated because the UV completion of
this model is a $USp(2N_c)$ gauge theory with an antisymmetric matter field. Such a matter content does not induce further zero modes allowing the generation of (\ref{monW}).

We then study the $R$ charges and the global charges of the monopoles to infer the 
constraints on the real masses induced by the  superpotential (\ref{monW}). 
For the R-charges we have
 \begin{equation}
\Delta_{T_0} = \Delta_{{\widetilde T}_0} = 2(1-\Delta_Q) + 2(1-\Delta_{\tilde Q}) -\Delta_X (N_c-1)
\end{equation}
Similar relations can be written down for the other global charges.
By imposing $\Delta_{T_0} =  \Delta_{\widetilde T_0}  = 2$ 
(and $Q^{(i)}_{T_0}= Q^{(i)}_{\widetilde T_0} = 0$ for the other global symmetries)
we observe that the constraint (\ref{b2}) is recovered .
This analysis can be performed in all of the dual phases, leading to the same constraint
 (\ref{b2}).
It has been shown in \cite{VdB} that also in this case  $W(D_6)$ is a symmetry of the three sphere partition function. We can also count the number of dual phases: there are $|W(D_6)|/| W(A_3) |^2 = \frac{2^5 6!}{(4!)^2}=40$ dual phases; again this corresponds to the calculation performed in \cite{Dimofte:2012pd}.

We now have two different theories in which $W(D_6)$ acts as a symmetry on the mass parameters.
It is tempting to conjecture a duality among such theories.
Such a duality can be proven at the level of the partition function.

 %
%
%
%
\tocless\subsubsection{$USp(2N_c)/U(N_c)$ dualities from the partition function}
%
%
%
%

Let us apply the Higgs and real mass flow just discussed  
on the LHS of  (\ref{E7}), performing the large $M$ limit on the various terms in both sides of the duality encoded in  (\ref{E7}).
We can simplify the calculation as explained in \cite{Benini:2017dud} by using the symmetries of the integrand. This is necessary also to obtain the correct dimension of the Weyl group when flowing from $USp(2N_c)$ to $U(N_c)$. 

Then we need to compare the divergent terms and only if they coincide we can read the equivalence between the finite parts. 
We will now just compute their phases by using 
 (\ref{iout}). 
 In the electric theory there are three sources of divergences,
 the contribution from the fundamental quarks,
the contribution from the antisymmetric field
and the contribution from the gauge sector.
%
After using the balancing condition (\ref{balancing}) the contribution $\sum_{i=1}^{N_c} \sigma_i$ vanishes. 
It signals the absence of an FI in the final result, as expected. This 
is consistent with the presence of the superpotential (\ref{monW}).
The phase is 
\begin{equation}
\label{phaseele}
\phi_e = 
-4 M N_c(2 \omega+\tau(N_c-1))
-2 N_c \omega  \sum _{i=1}^{4} (m _r-n _r)
+N_c \sum _{r=1}^{4} (m _r^2- n_r^2)
\end{equation}

Next we need to study the phase corresponding to the divergent pre-factor 
in the partition function of the dual theory.
In this case there is no Higgsing taking place and 
we have just two sources generating the large shift in $M$, the dual quarks and the mesons.
The dual quarks generating the shift are the one parameterized by $\nu_4 + \zeta$ and $\mu_4-\zeta$.
Their contribution to the phase 
together with the contribution of the  
massive mesons 
cancels 
(\ref{phaseele})
after imposing the balancing condition (\ref{balancing}).

Summarizing: all the phases cancel and one remains with the identity of {\bf Theorem 5.6.15} in \cite{VdB} with the same constraints
on the masses and the same duality map.
The identity is
\begin{eqnarray}
\label{finalD6}
Z_{U(N_c)_0} (\vec m;\vec n;\tau) 
=
\prod_{j=0}^{N_c-1} \prod_{r=1}^{3} \Gamma_h (j \tau + m_r + n_4,j \tau + m_4 + n_r)
\,
Z_{USp(2N_c)_0} (\vec \mu_\zeta;\tau) \end{eqnarray}
where 
\begin{equation}
\label{muzeta}
\mu_{\bf \zeta} = \{m_1+\zeta,m_2+\zeta,m_3+\zeta,n_1 - \zeta,n_2-\zeta,n_3-\zeta\}
\end{equation}
There is a constraint 
\begin{equation}
\label{eq:zeta2}
2 \zeta = m_4 + n_1 + n_2 +n_3 - 2 \omega + (N_c-1)\tau 
= 
2 \omega-(N_c-1)\tau -n_4 -m_1-m_2 -m_3
\end{equation}

This shows that one can obtain the duality between the $U(N_c)$ 
and the $USp(2N_c)$ theory discussed above
flowing from the $W(E_7)$ invariant case on the partition function.
The electric theory has superpotential (\ref{monW}), 
coming from the Higgs flow from $USp(2 N_c)$ to $U(N_c)$.

In the $USp(2N_c)$ dual theory the monopole superpotential is set to zero because of 
the real mass flow and we are dealing with a set of unconstrained real masses.
The correct constraints are imposed by the dual superpotential, 
involving the gauge singlets, identified with some of the mesons of the $U(n)$ theory.
The superpotential of the $USp(2N_c)$  dual theory
\footnote{Observe that this superpotential has already appeared in the literature in \cite{Amariti:2014lla}.
The duality discussed here can indeed be reduced to the one conjectured in  \cite{Amariti:2014lla} through a real mass flow.}
corresponds to the one discussed in \cite{Razamat:2017hda} and it is
\begin{equation}
\label{Wdd}
W = \sum_{j=1}^{N_c}  
 \tr\bigg(
M^{(j)} q q A^{j-1} 
+
N^{(j)} p p A^{j-1} 
\bigg)
\end{equation}
where $M$ and $N$ are gauge singlets, $q$ and $p$ are the quarks of the two $SU(3)$
global symmetries, and $A$ is in the antisymmetric of $USp(2N_c)$. 
The real masses of the quarks $q$ correspond
to the first three entries in (\ref{muzeta})
while the ones of the quarks $p$ to the last three entries.
The duality then identifies the singlets $M^{(j)}$ and $N^{(j)}$ with the mesons $Q_r \tilde Q_4 X^{N_c-j}$
and $Q_4 \tilde Q_r X^{N_c-j}$ respectively,
where $r=1,2,3$ and $j=1,\dots,N_c$. 
This imposes the constraints on $\zeta$ discussed in (\ref{eq:zeta2}),
breaking a baryonic-like symmetry.
It is interesting to observe that while the monopole superpotential of the electric theory (\ref{monW})
breaks the topological symmetry, the superpotential (\ref{Wdd}) of the dual theory, involving the mesonic
operators, breaks the baryonic symmetry. 
This behavior is reminiscent of mirror symmetry and may play a crucial role in a deeper understanding of this duality.

%
%
%
%
%
\subsection{The general scheme}
\label{subsec:general-scheme}
%
%
%
%
%
%
%
\begin{figure}
\begin{center}
\includegraphics[width=\textwidth]{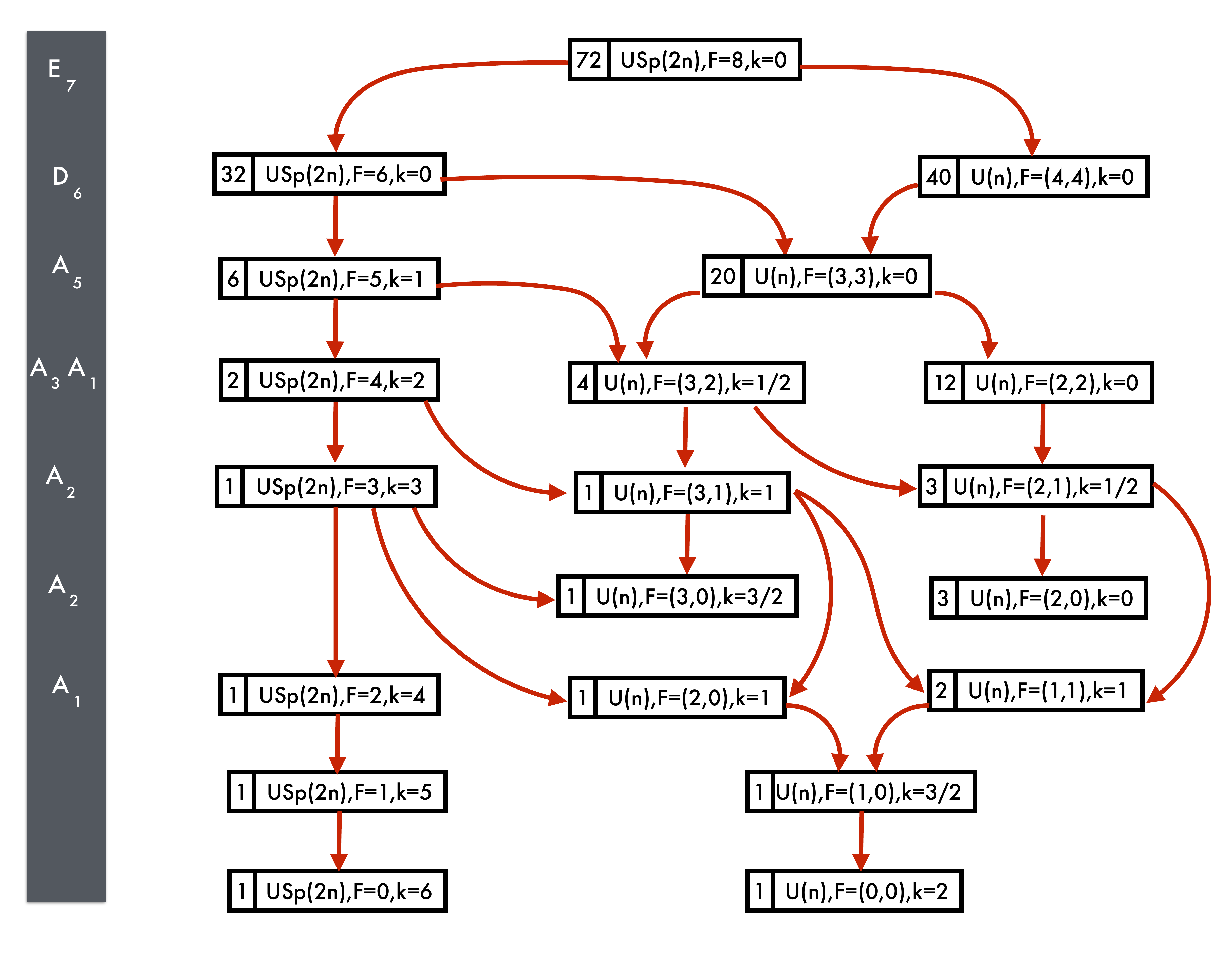}
\caption{Weyl group symmetry enhancements obtained from $USp(2N_c)$ with eight fundamentals
and antisymmetric defined on $\mathbb{R}^3 \times S^1$. 
The rectangles represent the partition function of sets of 3d $\mathcal{N}=2$ models,
with gauge group $USp(2N_c)_{2k}$ or $U(N_c)_{-k}$,  where $k$ refers to the CS level.
In the symplectic case there is one antisymmetric matter field $A$ and $F$ fundamentals, while in the unitary case there is one adjoint and a pair $F=(F_1,F_2)$ of fundamentals and anti-fundamentals..
The numbers appearing in a square on the left of each 
box correspond to the degeneration of integrands with the same 
gauge and charged matter content (up to spacetime parity and charge conjugation). 
They can in principle differ by the presence of hyperbolic gamma functions corresponding to 
extra singlets, i.e. that do not appear in the integrands.
These degenerations are obtained by modding the enhanced Weyl group symmetry of the integrands, 
defined in the grey column on the left of the picture, by
the manifest Weyl group symmetry, specified by the value(s) of $F$.
The red arrows in the figure specify the RG flow connecting the UV to the IR models in the diagram.
}
\label{fig2}
\end{center}
\end{figure}

In this section we study the general web of dualities and enhancement of the 
Weyl group symmetry that can be derived from the reduction of 4d $\mathcal{N}=1$  
$USp(2N_c)$ with eight fundamentals and one antisymmetric.
This web is obtained generalizing the real mass and Higgs flows that led
to the dual models with the $W(D_6)$ enhancement.
In order to do that let's discuss formally the flow from the case with $W(E_7)$ enhancement to the case with $W(D_6)$ enhancement.

In that case we reduced the manifest global symmetry by assigning real masses. 
The partition function had a reduced discrete symmetry group, corresponding to 
a $W(D_6)$ subgroup of $W(E_7)$. 
There are different ways to construct such subgroups,
corresponding to new dualities, transforming for example $USp(2N_c)$ into $U(N_c)$.

The discussion was made mathematically precise in \cite{VdB} by 
showing that the symmetry of the new integrals is a 
 subgroup of $W(E_7)$.
This  subgroup acts as a discrete symmetry on the IR partition functions
and it is obtained by letting the
variables $\mu$ of the $E_7$ case go to infinity in the direction of a vector in eight dimensional Euclidean space.
This vector corresponds to 
$\vec b_1=(0,0,0,0,0,0,1,-1)$ in the first case and to
$\vec b_2= (1,1,1,1,-1,-1,-1,-1)$ in the second case (up to normalizations).
Observe that the vectors $\vec b_i$  are defined modulo permutations. These permutations 
explore the degenerations of the 32 $USp(2N_c)$ models with six fundamentals
and of the 40 $U(N_c)$ models with four flavors.
The vectors orthogonal to $\vec  b_i$ form two different embeddings of 
the root system of $D_6$.
The discrete symmetry of the IR partition functions
corresponds to the reflections in these roots.
It is possible to transform the system defined by $\vec b_1$ to the
system defined by  $\vec b_2$
by acting with the 
``broken'' elements of $W(E_7)$: this  is the mathematical interpretation of the 
$USp(2N_c)/U(N_c)$ duality that we discussed above at physical level.
The classification scheme has been completely carried out in \cite{VdB} and here we  report 
the results, translating them in a physical language.

In order to classify the other possible Weyl group symmetry enhancements we need to 
iterate the flow, by further real mass and Higgs flows.
In this way we can construct models with $USp(2N_c)_{2 \kappa}$ 
and $U(N_c)_{\kappa}$ gauge groups, antisymmetric matter and a lower amount of 
fundamentals.
These flows preserve the dualities and this translates in a possible enhancement of the Weyl group symmetry
for some of the IR theories.
For example starting from the $W(D_6)$ case one can flow to 
a case with enhancement of the Weyl group symmetry to $W(A_5)$.
In terms of the gauge group and of the charged matter content we have two possibilities (counted up to parity transformations).

\begin{itemize}
\item $USp(2N_c)_{1}$ with five fundamentals and an antisymmetric.
The Weyl group symmetry in this case enhances to  $W(A_5)$. Accordingly, there are 
$\frac{|W(A_5)|}{|W(A_4)|} = 6$ dual models.
\item $U(N_c)_0$ with three fundamental flavors and an adjoint. Also in this case
the Weyl group symmetry enhances to  $W(A_5)$
and there are $\frac{|W(A_5)|}{|W(A_2)|^2} = 20 $ models.
\end{itemize}

In Figure \ref{fig2} we reported the full structure of the RG flow, by iterating the procedure.
At each level we specify the degeneration of the models, and we can observe that is it always consistent 
with the ratio of the orders of the enhanced Weyl group and that of the Weyl group of the naive global symmetry of the classical action.

From the figure we can extract some physical consequences.

Each row represents a set of models with the same three sphere partition function.
This is the first step to claim a duality between these models.
The equivalences among the various partition functions hold if 
the correct duality maps, the CS contact terms and the balancing conditions
are specified.
These constraints can be obtained by studying the flows from the UV models. 
We refer the interested reader to \cite{VdB} where these results have been computed
and listed.

Moreover each box in the figure is associated to a degeneration of integrals, and consequently of models.
The manifest global symmetry in each of these models can enhance to a larger symmetry group.
This can happen by consistently deforming the superpotentials, in the various dual phases,
transforming the duality into a self duality. In this way the Weyl group symmetry enhancement 
of the integrands becomes a discrete symmetry enhancement of the full partition functions.
This is a necessary condition for the global symmetry enhancement. More refined analyses (for example
the analysis of the operators counted by the superconformal index, or the study of the Hilbert series) 
are then necessary in order to see if the Weyl group symmetry enhancement can be promoted to an enhancement of global symmetry to the full group/algebra.

We conclude with an observation about the $U(N_c)_0$ model with two fundamentals and an adjoint.
In this case one can observe an enhancement of the Weyl group symmetry to $W(A_2)$.
In the limit of $N_c=1$ this model coincides
with the model discussed in 
 \cite{Gang:2018wek,Gaiotto:2018yjh,Benini:2018bhk},
where the global symmetry has been indeed conjectured to enhance to $SU(3)$.
Here we observe
\footnote{Furthermore in \cite{Fazzi:2018rkr} it has been shown that
this model is dual to  an
$SU(3)_{5/2}$ gauge theory with 
a manifest $SU(3)$ global symmetry.
}, at the level of the partition function,
that the model is also dual to a  $U(1)_{-3/2}$
gauge theory with three fields at charge 1 (see also \cite{Fazzi:2018rkr}).
This last theory has a non trivial monopole superpotential
and it corresponds to a self dual case
of the duality studied in \cite{Benini:2017dud} with $U(N_c)_{k/2}$ gauge groups,
$N_f$ fundamentals and $N_f-k$ anti-fundamentals.

Finally, we also observe that 
the enhancement of $U(N_c)$ with two pairs of fundamental flavors
to $A_3 \times A_1$ corresponds to the $SO(6)$ enhancement discussed in
\cite{Benini:2018bhk}.

%
%
%
%
\section{Six fundamentals and confining theories}
\label{sec:confining}
%
%
%
%
In this section we study the dimensional reduction of a $USp(2N_c)$ gauge theory with 
an antisymmetric and six fundamentals.
This model can be obtained from the one with eight fundamentals by 
a superpotential mass deformation.
It has been observed in \cite{Spiridonov:2008zr} that in this case the superconformal index 
supports an enhancement of the global symmetry to $E_6$
\footnote{See also \cite{Razamat:2017wsk} for related discussions.}.
Here we will not comment on the enhancement of the global symmetry 
for this case.
We will rather study the consequences of this mass deformation in the dual 
model with the maximal amount of global symmetry, which becomes a confined
WZ model.
The 4d theory was studied in \cite{Csaki:1996eu}, and it was indeed shown that,
in the IR, this theory confines without breaking the chiral symmetry.
The confined degrees of freedom are expressed in terms of gauge invariant combination 
of the matter fields. They correspond to the gauge invariant operators
\begin{eqnarray}
&&
S_k  \propto \tr\, A^{k+2},
\,
M_k \propto Q A^k Q,
\quad k=0,1,\dots,N_c-1
\end{eqnarray}
There is also a superpotential interaction in the confined description, with a 
number of terms rapidly growing with $N_c$.
For  $N_c=2,3,4$ these superpotentials have been given in \cite{Csaki:1996eu}.

This theory can be reduced to 3d, in both the  confining and in the confined phase.
In the first case one has a 3d effective $USp(2N_c)$ gauge theory with 
an antisymmetric and six fundamentals. This theory has also a KK monopole 
superpotential, $W = \eta Y$.
The confined theory on the other hand has the same fields and interaction of the 4d parent.
This is consistent with the results of \cite{Csaki:2014cwa,Amariti:2015kha}, where the reduction of $U(N_c)$ confining gauge theories was discussed in details.

The 3d duality obtained by this reduction has a well studied mathematical counterpart 
in the analysis of hyperbolic integrals.
Indeed the matching between the partition function relating the two theories 
was already proven by \cite{VanDiejen:2001,VanDiejen:2005,Rains:2006}.
The explicit relation is 
\begin{eqnarray}
Z_{USp(2N_c)}(\vec \mu;\tau)
=
\prod_{j=2}^{N_c} \Gamma_h(j \tau) \prod_{j=0}^{N_c-1} 
\prod_{1\leq a<b \leq6}\Gamma_h(j \tau+\mu_a +\mu_b)
\end{eqnarray}
with the relation among the mass parameters
\begin{equation}
2(N_c-1) \tau +\sum_{a=1}^6 \mu_a = 2 \omega
\end{equation}
This relation, that corresponds to impose the anomaly free constraints
on the 4d R-current, signals the presence of the  KK monopole superpotential in the confining theory. In the confined case 
it is consistent with the superpotentials of \cite{Csaki:1996eu}.
\\
\\
It is possible to study a 3d conventional limit by real mass and
Higgs flow. We consider these two cases separately in the following sections.  

\subsection{Real mass flow}

In this case we assign a large mass to two fundamentals reducing the theory to 
$USp(2N_c)$ with four fundamentals and an antisymmetric.
In such cases the  meson $M_{ij}$, in the antisymmetric representation of $SU(6)$
splits into an antisymmetric meson of $SU(4)$ and a monopole.
For example, if we consider the case $N_c=2$, the superpotential of the confined
phase, before turning on the real mass, is
\begin{equation}
W 
= S_2 \epsilon_{ijklmn} M_{0}^{ij} M_{0}^{kl} M_{0}^{mn} 
+
\frac{1}{3}\epsilon_{ijklmn} M_{0}^{ij} M_{1}^{kl} M_{1}^{mn} 
 \end{equation}
In the IR we have
\begin{equation}
\label{Wmonred}
W 
= S_2 Y_0 \epsilon_{ijkl} M_{0}^{ij} M_{0}^{kl}  
+
\frac{1}{3} Y_0 \epsilon_{ijkl} M_{1}^{ij} M_{1}^{kl} 
+
\frac{1}{3} Y_1 \epsilon_{ijkl} M_{0}^{ij} M_{1}^{kl} 
\end{equation}
where the fields $Y_i = Y\,\tr\,A^i$
corresponds to the dressed monopoles of the confining theory
acting as singlets in the confined phase.
It can be indeed checked that the charges
of $Y_i$ obtained from the superpotential (\ref{Wmonred})
correspond to the ones obtained by using the quantum formula for
the monopole charge (\ref{monch}) in the electric theory.
This claim can be corroborated by studying the partition function.
Indeed in this case we arrive at the identity
\begin{eqnarray}
\label{Zeq4}
Z_{USp(2N_c)}(\vec \mu;\tau)
&=&
 \prod_{j=0}^{N_c-1} 
\Gamma_h \Big(2 \omega - (2N_c-2-j)\tau-\sum_{a=1}^4 \mu_a \Big)
\nonumber 
\\
&\times &
\prod_{j=2}^{N_c} \Gamma_h(j \tau)
\prod_{1\leq a<b \leq 4}\Gamma_h(j \tau+\mu_a +\mu_b)
\end{eqnarray}
where the mass parameters are unconstrained.
The first term in the RHS of (\ref{Zeq4})
corresponds to the contribution of the dressed monopoles
$Y_i$, for $i=0,\dots,N_c-1$. 

\subsection{Higgs flow}

A second interesting 3d limit can be taken by shifting the scalars $\sigma_i$ 
by a large real quantity $s$. If this flow is supported by a real mass flow $\mu_a \rightarrow m_a+ s$
and $\mu_{a +3}\rightarrow n_a- s$ for $a=1,2,3$, the final theory has $U(N_c)$
gauge group, three pairs of fundamental and antifundamental quarks and a monopole 
superpotential
\begin{equation}
W = T_0 + \widetilde{T}_0
\end{equation}
This theory is dual to a set of singlets interacting through a superpotential. For example if $n=2$ we have
\begin{equation}
W = S_2  \epsilon_{ijk} \epsilon_{lmn} M_{0}^{il} M_{0}^{jm}  M_{0}^{kn}  
  + \frac{1}{3} \epsilon_{ijk} \epsilon_{lmn} M_{0}^{il} M_{1}^{jm}  M_{1}^{kn}  
\end{equation}
where the mesons $M_j^{ab} = Q_a X^j \tilde Q_b$ are in the bifundamental 
representation of the $SU(3)$ non-abelian flavor symmetry group.
Here $X$ represents the adjoint matter field and $j=0,\dots,N_c-1$.
In this case there are no monopoles of the electric theory acting as singlets in the magnetic dual. Indeed all the massless components of the original (antisymmetric) mesons become components of the new (bifundamental) meson in the theory with the reduced flavor.
It is possible to reproduce this behavior on the partition function. Indeed this duality corresponds to an exact identity obtained in \cite{VdB}.
The identity is
\begin{eqnarray}
\label{Zeq5}
Z_{U(n)_0}(\vec m;\vec n;\tau) =
\prod_{j=2}^{N_c} \Gamma_h(j \tau) \prod_{j=0}^{N_c-1} 
\prod_{a,b=1}^{ 4}\Gamma_h(j \tau+m_a +n_b)
\end{eqnarray}
where the parameters satisfy the relation
\begin{equation}
2(N_c-1)\tau + \sum_{a=1}^{3} (m_a+n_a) = 2 \omega
\end{equation}
Observe that the symmetries of the integrals in (\ref{Zeq4}) and (\ref{Zeq5}) are consistent with the $D_5$ enhancement 
\cite{Gahramanov:2013xsa}. It would be interesting to further study this enhancement along the lines of section \ref{sec:E7}.
%
%
%
%
%
%
%
\section{Conclusions}
\label{sec:conc}
%
%
%
%
%
%
%
In this paper we reduced 4d theories with antisymmetric matter fields to 3d, finding new dualities and 
relations.
We   summarize here our main results.
\begin{itemize}
\item In section \ref{sec:SPas}  we completed the picture in Figure \ref{fig1}, showing that 
the general aspects of the reduction of $USp(2N_c)$/$U(N_c)$ dualities with fundamental matter persist
when adding anti-symmetric/adjoint matter with a power law superpotential.
We have provided arguments from field theory and localization to confirm our claims.
\item
In section \ref{sec:E7} we obtained a family of 3d effective $USp(2N_c)$ theory with eight fundamentals and one anti-symmetric matter field in which the action of the $E_7$ Weyl group is manifest on the real masses, leading to the generalization of the dualities of 
\cite{Spiridonov:2008zr}
to 3d.
Furthermore we constructed a whole web of $USp(2N_c)/U(N_c)$ dualities, generalizing the $SU(2)/U(1)$ results of \cite{Dimofte:2012pd} to higher ranks and to lower symmetry groups, as summarized in Figure \ref{fig2}.
\item
In section \ref{sec:confining} we reduced 4d confining 
$USp(2N_c)$ theories with six fundamentals and an antisymmetric matter field.
Also in this case we obtained new relations for both $USp(2N_c)$ and for $U(N_c)$ theories.
\end{itemize}

There are some  general lessons that we can extract from our analysis.
One of  them regards the structure of the monopole superpotentials in the 
presence of unitary gauge groups and adjoint matter.
If such a theory is considered as UV complete the zero modes counting does not allow the generation 
of any monopole superpotential, because each adjoint field carries two zero modes.
Nevertheless, as we widely discussed in the paper, we claimed the existence of linear monopole/anti-monopole superpotentials. Such terms correspond to monopoles of the UV completion, i.e. 
$USp(2N_c)$ models with anti-symmetric matter, where the zero mode counting allows the
generation of the monopole superpotentials. 
 It would be interesting to further investigate this phenomenon, finding other example of
its applicability and discussing its relation with the index theorems that in general prevent the generation
of the superpotentials that we have constructed here. 

The problem of finding an UV completion should be analyzed together with the study of accidental symmetries.
This requires the minimization of the partition function 
\cite{Jafferis:2010un} and 
an analysis along the lines of 
\cite{Morita:2011cs,Agarwal:2012wd,Safdi:2012re}.
We leave this interesting problem to future analysis.

Another result that requires some investigation is the extension of the flows discussed in Figure \ref{fig1} to the
case of Brodie duality \cite{Brodie:1996vx}, involving two adjoints and a non trivial superpotential.
Recently, this duality has been reduced to 3d in \cite{Hwang:2018uyj}. The main difference that emerged from the analisys is the presence of superpotentials involving monopoles with charge 2.
It would be interesting to study how this behavior modifies our analysis and if a structure similar to the one 
in Figure \ref{fig1} does appear for this case as well.

Further studies may involve the existence of 3d dualities between 
$USp(2N_c)$ and $U(N_c)$ theories. It should be interesting to investigate if and under which conditions one can find 4d analogs of these dualities. Another interesting possibility regards the existence of 3d dualities between orthogonal and unitary gauge theories.

We wish to conclude with a comment on the global symmetry enhancements: here we have observed the
fact that, thanks to localization, it is possible to prove that the action of the Weyl group is larger that the
one expected from the classical global symmetry.
It is by itself a signal of the possibility of a global symmetry enhancement but further investigations are
usually necessary for a complete understanding. Anyway we think that this mechanism can be used as a guideline for the existence of this sort of enhancements.

\acknowledgments
We thank D.~Orlando, S.~Reffert and S.~Penati for valuable discussions and early collaboration
on this project.
We are also grateful to Sergio Benvenuti for informing us of its work \cite{Benvenuti:2018bav}.
We thank Marco Fazzi for comments on the draft and on related issues.
This work is supported in part by Italian Ministero dell'Istruzione, Universit\`a e Ricerca (MIUR) and Istituto Nazionale di Fisica Nucleare (INFN) through the ``Gauge Theories, Strings, Supergravity'' (GSS) research project.
A.A. thanks the Galileo Galilei Institute for Theoretical Physics for the hospitality and the INFN for partial support during the completion of this work.
%
%
%
%
%
%
%
\appendix
%
%
%
%
%

%
%
%
%
%
\section{Aspects of 3d $\mathcal{N}=2$ theories}
\label{sec:app}
%
%
%
%
%
%
%
In this appendix we collect some  relevant results on 3d $\mathcal{N}=2$ theories.
We focus on the reduction of 4d dualities to 3d dualities and its realization in terms of localization.

\subsection{4d/3d reduction and KK monopole}
\label{APPA}
Preserving a 4d supersymmetric duality in 3d can be done by compactifying the dual phases 
on a finite size circle. The procedure consists of dimensionally reduce the field content
 and to add the effective 3d dynamics due to the finite size.
This lifts possible 4d anomalous symmetries that can potentially become non-anomalous in 3d.
Such symmetries are indeed broken by the presence of superpotential terms involving the
KK monopoles. The KK monopoles contribute to the effective superpotential if in the spectrum 
there are only two fermionic zero modes, coming from the gaugino in the adjoint representation, 
while the matter fields do not carry further fermionic zero modes
\footnote{Actually this condition can be made milder in the presence of potential interactions
involving the fermions carrying the extra zero modes.}.
The counting of these zero modes follows from an application of the index theorem. 
Essentially the circle compactification splits the 4d instanton in a set of
BPS monopoles, counted by the Callias theorem \cite{Callias:1977kg}, and in one KK monopole. 
The total amount of zero modes for these configurations corresponds to the number of
zero modes of the original 4d instanton, obtained from the Atiyah--Singer index
theorem. The difference between the two indices counts the number of zero modes 
in the KK monopole.
A more direct result follows from \cite{Poppitz:2008hr}, where an index theorem on $\mathbb{R}^3 \times S^1$ was derived (see also \cite{Nye:2000eg}).
In this case the counting of the zero modes in the KK monopole background
from each matter fields associated to the affine root.
For example the presence of fundamental matter fields does not
modify the number of zero modes and the KK monopole is generated. 
Here we have been interested in $USp(2N_c)$ gauge groups with antisymmetric matter. 
The index of \cite{Poppitz:2008hr} has been computed for this representation in \cite{Golkar:2009aq}, and one can
see that also in this case the KK monopole superpotential is generated, because no further 
fermionic zero modes associated to the affine root are present.\\
%
%
%
%
%
\subsection{Counting of zero modes}
\label{app:counting}
%
%
%
%
%
%
%
In this section we review the counting of zero modes of fermions in a
monopole/instanton background. Our derivation will closely follow that of
\cite{Callias:1977kg} and \cite{Poppitz:2008hr}.
%
%
%
%
%
%
%
\tocless\subsubsection{Lie algebra conventions}
\label{sec:liealg}
%
%
%
%
%
%
%
Here we collect a few well known facts about root systems of Lie algebras
which are useful in the counting of zero modes. For a standard reference on
the subject see \cite{Humphreys:1972}.

Let $G$ be a connected, simply connected and semisimple Lie group with Lie
algebra $\mathfrak{g}$. The rank $r$ of $G$ is the dimension of any of its
maximal torus subgroups or equivalently the dimension of the corresponding
Cartan subalgebra. Having chosen a particular such Cartan subalgebra
$\mathfrak{h} \subset \mathfrak{g}$ we pick a basis of commuting generators
$\{ h_i \}_{i = 1}^r \subset \mathfrak{h}$ that satisfy:
\begin{equation}
  \tr_{\mathcal{R}} [h_i h_j] = T (\mathcal{R}) \delta_{i j}
\end{equation}
for any irreducible representation $\mathcal{R}$ of $\mathfrak{g}$.

We denote the roots of the algebra as $\Delta = \{ \alpha_i \}_{i = 1}^{\dim
(G) - r} \subset \mathfrak{h}^{\ast}$, the simple roots as $\{ \beta_i \}_{i =
1}^r \subset \Delta$ and the inverse of any root $\alpha$ in the
root system $\Delta$ as:
\begin{equation}
  \begin{array}{ccc}
    \alpha^{\vee} = \frac{2 \alpha}{\alpha \cdot \alpha} \in
    \mathfrak{h}^{\ast} & \hspace{3em} & \alpha \cdot \alpha \equiv \sum_{i =
    1}^r \alpha (h_i) \alpha (h_i)
  \end{array}
\end{equation}
We define the co-roots $H_i$ to be the duals of the inverse simple roots $\beta^\vee_j$, i.e., those elements of the cartan algebra that satisfy the relation:
\begin{equation}
 w(H_i)=\beta^\vee_i\cdot w
\end{equation}
for any weight $w\in\mathfrak{h}^\ast$ of the algebra.

A particular choice of simple roots defines an associated fundamental Weyl
chamber corresponding to the convex subset $\{ v \in \mathfrak{h} \,|\, \beta_i
(v) > 0, \forall i = 1, \ldots, r \} \subset \mathfrak{h}$.
Moreover one can split the root system $\Delta$ into two components:
\begin{equation}
  \Delta = \Delta^+ \cup \Delta^-
\end{equation}
where $\Delta^+$ ($\Delta^-$) are the positive (negative) roots, i.e., those
that are positive (negative) integer combinations of the simple roots.
The Weyl vector $\rho$ is then defined as the half-sum of all the positive
roots:
\begin{equation}
  \rho = \frac{1}{2} \sum_{\alpha \in \Delta^+} \alpha \label{eq:weylvect}
\end{equation}
Given a root system $\Delta$ and a choice of simple roots $\{ \beta_i \}$ one
can define a partial order on $\Delta^+$ as follows. For any positive root
$\alpha = \sum_{i = 1}^r m_i \beta_i$, define the degree (or level) of
$\alpha$ as:
\begin{equation}
  \deg (\alpha) = \sum_{i = 1}^r m_i \in \mathbb{Z}
\end{equation}
then the degree map endowes $\Delta^+$ with the structure of a partially
ordered set, the root poset. The highest root is the root with the highest
degree and it is unique with respect to this property. It is customary to
write the highest root and its inverse as:
\begin{equation}
  \theta = \sum_{i = 1}^r k_i \beta_i \quad\text{and}\quad
  \theta^\vee = \sum_{i = 1}^r k_i^\vee \beta_i^\vee \label{eq:highestroot}
\end{equation}
where $k_i$ are called the Kac labels of the algebra and $k_i^\vee$ are the Dynkin numbers. We will refer to the lowest root $-\theta$ as the affine root and define its dual co-root $H_0$ as:
\begin{equation}
  H_0=-k_i^\vee H_i
  \label{eq:affinecoroot}
\end{equation}
so that $w(H_0)=-\theta^\vee\cdot w$ for every weight $w$.\\

The weights of $G$ form a lattice in $\mathfrak{h}^{\ast}$ generated by the fundamental weights $\{\lambda_i\}$ defined by the relation:
\begin{equation}
  \beta^{\vee}_i \cdot \lambda_j = \lambda_j(H_i) = \delta_{i j}
  \label{eq:fundamentalweights}
\end{equation}
By definition then the weight lattice and the co-root lattice are integral dual to each other.
It is a well known result that the sum of all fundamental weights coincides
with the Weyl vector:
\begin{equation}
  \rho = \sum_{i = 1}^r \lambda_i
  \label{eq:weylvector}
\end{equation}
which then implies that $\rho(H_i)=\beta^{\vee}_i \cdot \rho = 1$ for every co-root $H_i$.\\

Finally, we recall the useful formula:
\begin{equation}
  \Lambda \cdot (\Lambda + 2 \rho) = C_2 (\mathcal{R})
  \label{eq:casimir}
\end{equation}
where $\Lambda$ is the highest weight of the representation
$\mathcal{R}$ and $C_2 (\mathcal{R})$ is the value of the
quadratic Casimir element in that representation.

\tocless\subsubsection{Callias index theorem}
\label{app:callias}

Consider a Euclidean theory on $\mathbb{R}^3 \times S^1$ with a massless
Dirac fermion $\psi$ in the representation $\mathcal{R}$ of the gauge group
$G$. Coordinates are chosen as $\{ x^i \}_{i = 1, 2, 3}$ on $\mathbb{R}^3$
and $x^4$ on $S^1$.
We look for solutions of the Dirac equation for $\psi$, i.e., zero
eigenfunctions of the Dirac operator:
\begin{equation}
  D \psi = \gamma^{\mu} (\partial_{\mu} + A_\mu) \psi = \left[
  \begin{array}{cc}
    0 & - \fredholm^{\dagger}\\
    \fredholm & 0
  \end{array} \right] \left[ \begin{array}{c}
    \psi^+\\
    \psi^-
  \end{array} \right] = 0
\end{equation}
where $\fredholm$ and $\fredholm^{\dagger}$ are Fredholm operators acting on spinors of
definite chirality and, on an anti-selfdual background, satisfy:
\begin{equation}
  \fredholm^{\dagger}\, \fredholm = - D_{\mu} D^{\mu} + 2 \gamma^m B_m \quad \mathrm{and}
  \quad \fredholm\, \fredholm^{\dagger} = - D_{\mu} D^{\mu}
\end{equation}
where $B_m = \frac{1}{2} \varepsilon_{m l k} F_{l k} = F_{4 m}$ is the magnetic field on $\mathbb{R}^3$.\\

What we are interested in computing is the difference in the number of zero
modes of $\fredholm$ and those of $\fredholm^{\dagger}$. This quantity is a topological
invariant and is called the index of $\fredholm$:
\begin{equation}
  I_{\mathcal{R}} \equiv \ind(\fredholm) = \dim \ker (\fredholm) - \dim \ker (\fredholm^{\dagger})
  \label{eq:indexL}
\end{equation}
Using the fact that $\ker (\fredholm^{\dagger}\, \fredholm) = \ker (\fredholm)$ and
$\ker (\fredholm\, \fredholm^{\dagger}) = \ker (\fredholm^{\dagger})$, the index
can be conveniently computed by the formula:
\begin{equation}
  I_{\mathcal{R}} = \lim_{M^2 \rightarrow 0} \tr_{\mathcal{R}} \left[
  \frac{M^2}{\fredholm^{\dagger}\, \fredholm + M^2} \right] - \tr_{\mathcal{R}} \left[
  \frac{M^2}{\fredholm\, \fredholm^{\dagger} + M^2} \right] \label{eq:index}
\end{equation}
Observe that the trace in (\ref{eq:index}) is both over the representation
$\mathcal{R}$ and over the Hilbert space on which the differential operator
$\gamma^{\mu} \partial_{\mu}$ acts, i.e., the Hilbert space of sections of the
spinor bundle.

In terms of the 4d Dirac operator we can write:
\begin{equation}
  I_{\mathcal{R}} (M^2) = \tr_{\mathcal{R}} \left[ \gamma_5 
  \frac{M^2}{- D^2 + M^2} \right] = M \tr_{\mathcal{R}} \left[ \gamma_5 
  \frac{D + M}{- D^2 + M^2} \right] = M \tr_{\mathcal{R}} \left[
  \gamma_5  \frac{1}{- D + M} \right]
\end{equation}
Observe that in the previous formula it appears the propagator of the Dirac
fermion $\psi$ as:
\begin{equation}
  \langle \psi (x_1) \bar{\psi} (x_2) \rangle = \left\langle x_1 \left|
  \frac{1}{D - M} \right| x_2 \right\rangle
\end{equation}
corresponding to the Euclidean action $- \bar{\psi} (- D + M) \psi$ where the
mass $M$ has been introduced as an auxiliary parameter. Hence we can write:
\begin{equation}
  I_{\mathcal{R}} (M^2) = - M \tr_{\mathcal{R}} [\gamma_5  \langle \psi
  \bar{\psi} \rangle] = \int_{S^1} \mathd x^4 \int_{\mathbb{R}^3} \mathd^3
  x\,M \tr_{\mathcal{R}} \langle \bar{\psi} \gamma_5 \psi
  \rangle \label{eq:indexM}
\end{equation}
The r.h.s. of (\ref{eq:indexM}) can be expressed using the Atiyah-Singer
index theorem for the abelian anomaly of a 4d theory with a massive Dirac
fermion $\psi$:
\begin{equation}
  \partial_{\mu} J^{\mu}_5 \equiv \partial_{\mu} (\bar{\psi} \gamma^{\mu}
  \gamma_5 \psi) = - 2 M \bar{\psi} \gamma_5 \psi - \mathrm{ch}_2 (F)
  \label{eq:current}
\end{equation}
where the second term on the right of (\ref{eq:current}) is the second Chern
character of the bundle associated to the representation $\mathcal{R}$:
\begin{equation}
  \mathrm{ch}_2 (F) = \frac{1}{8 \pi^2} \tr_{\mathcal{R}} [F
  \wedge F]
\end{equation}
Substituting (\ref{eq:current}) in (\ref{eq:indexM}) the index can finally be
rewritten as:
\begin{equation}
\label{eq:index2}
  I_{\mathcal{R}} (M^2) = - \frac{1}{2} \int_{S^1} \mathd x^4
  \int_{S^2_{\infty}} \varepsilon_{i j k}  \langle J^i_5
  \rangle \mathd x^j \mathd x^k - \frac{1}{2} \int_{\mathbb{R}^3
  \times S^1} \mathrm{ch}_2 (F)
\end{equation}
%
%
%
%
%
\tocless\subsubsection{BPS monopole background}
\label{app:bpsmonopole}
%
%
%
%
%
%
%
The first type of background for which we are interested in counting fermionic
zero modes is that of a static 4d monopole solution of 't
Hooft-Polyakov. For a more general gauge group $G$ we
consider the embedding of the $SU (2)$ solution into the group $G$ as
in \cite{Weinberg:1979zt,Weinberg:1982ev}. These are usually referred to as BPS monopole backgrounds.

Because the solution is ``static'', the fourth component $A_4$ of the gauge
connection behaves effectively as a Higgs field $\Phi$ for the connection
$A_i$ on $\mathbb{R}^3$. The theorem of Callias then states that the index of
(\ref{eq:indexL}) depends on the topology of the Higgs field by counting the
winding number of the map $| \Phi |^{-1}\Phi$ as it goes around the
2-sphere at spatial infinity.

In this set up we have that the Higgs field $\Phi \sim A_4$ is constant in
$x^4$ but varies along $\mathbb{R}^3$. Finiteness of the energy of the solution imposes the following restrictions:
\begin{equation}
 |\Phi| \rightarrow 1,\quad F\rightarrow 0,\quad D\Phi\rightarrow 0 \quad \text{for}\quad|x|\rightarrow \infty
\end{equation}
so that the connection is asymptotically pure gauge. Because, the Higgs field is covariantly constant on the sphere at spatial infinity, we can write:
\begin{equation}
\Phi(x)|_{S^2_\infty} = \mathrm{Ad}_{g(x)}\vev
\end{equation}
where $\vev\equiv\Phi(p)$ is the value of the Higgs field at some fixed reference point $p\in S^2_\infty$ and $g:S^2_\infty\rightarrow G$. This corresponds to a global trivialization of the adjoint bundle of which $\Phi$ is a section and can always be done since the bundle is topologically trivial on $\mathbb{R}^3$.

The field $\Phi$ is then equivalent to the pair $(\vev,g)$ and defines a map to the orbit of the VEV $\vev$ in the Lie algebra $\mathfrak{g}$ under the adjoint action of $G$.
We assume that the VEV breaks the gauge group maximally, i.e., we choose $\vev$ such that its stabilizer in $G$ is a maximal torus $T\cong U (1)^r \subset G$.
This implies that orbit is isomorphic the coset space $G/T$ and:
\begin{equation}
  \Phi : S^2_{\infty} \rightarrow G/T
\end{equation}
The homotopy class of the Higgs field then specifies an element of the second
homotopy group of this coset, which for $G$ simple and simply connected is:
\begin{equation}
  \pi_2 (G / T) \cong \mathbb{Z}^r
\end{equation}
With this choice of VEV the BPS solution can be written explicitly as:
\begin{equation}
  A_4 |_{S^2_{\infty}} = \vev \label{eq:defv}
\end{equation}
\begin{equation}
  \label{eq:defn}
  F |_{S^2_{\infty}} = \frac{\cochar }{2}
  \frac{\varepsilon_{ijk} x^i\mathd x^j\mathd x^k}{|x|^3} \quad \mathrm{with}
  \quad \cochar  \equiv \sum_{i=1}^r n_i H_i
\end{equation}
where we have performed a patch-wise gauge transformation to make the Higgs field constant.\\

The solution is described by the following parameters:
\begin{itemize}
  \item $\vev\in\mathfrak{h}$ parametrizes the choice of asymptotic VEV for the
  Higgs field $A_4$. We choose $\vev$ such that:
  \begin{equation}
    \beta_i (\vev) > 0
    \label{eq:weylchamber}
  \end{equation}
  for any simple root $\beta_i \in \mathfrak{h}^{\ast}$. It follows that
  $\vev$ is \textit{regular} with respect to the chosen basis of
  simple roots $\beta_i$ and lies into the \emph{fundamental Weyl chamber}
  (other choices are possible but are all related by the action of the Weyl
  group).
  
  \item $n_i \in \mathbb{Z}$ is the magnetic charge\footnote{For each fundamental
  monopole to be properly quantized as an $SU(2)$ BPS solution embedded into the
  larger gauge group $G$ we need $\cochar$ to be a vector in the co-root lattice of
  $G$. In fact, if $\cochar$ is an integer linear combination of co-roots then its
  eigenvalues are integers in all representations. This can be shown by
  considering that every weight $w$ is a linear combination of fundamental
  weights $\lambda_i$ with integer coefficients, so that, using
  (\ref{eq:fundamentalweights}) we have $w(\cochar) \in \mathbb{Z}$.} of the
  fundamental BPS monopole associated to the simple co-root $H_i \in \mathfrak{
  h}$; these integers describe the topology of the Higgs field and formally
  correspond to elements of $\pi_2 (G / T)$.
\end{itemize}

Substituting (\ref{eq:defv}) and (\ref{eq:defn}) in (\ref{eq:index2})
and taking the $M\to 0$ limit, the limit is given by \cite{Poppitz:2008hr}:
\begin{equation}
  I^{\mathrm{BPS}}_{\mathcal{R}} = \tr_{\mathcal{R}} \left[ \left\lfloor
  \frac{\vev  L}{2 \pi} \right\rfloor  \cochar  \right] = \sum_{i =
  1}^{\dim (\mathcal{R})} \left\lfloor w_i \left( \frac{\vev  L}{2 \pi}
  \right) \right\rfloor w_i (\cochar )
  \label{eq:bps}
\end{equation}
where $L$ is the size of the circle $S^1$, $\lfloor x \rfloor = \max \{ m\in \mathbb{Z} \, |\, m \leq x \}$ is the floor function
and $w_i$ are the weights of the representation $\mathcal{R}$.\\

In the small radius limit we have $|w_i\left(\tfrac{\vev L}{2\pi}\right)| \ll 1$ for every weight $w_i$ so that:
\begin{equation}
  \left\lfloor w_i \left( \frac{\vev  L}{2 \pi} \right)
  \right\rfloor = \left\{ \begin{array}{ccc}
    0 & \mathrm{for} & 0 \leq w_i \left( \frac{\vev  L}{2 \pi}
    \right) < 1\\
    - 1 & \mathrm{for} & - 1 < w_i \left( \frac{\vev  L}{2 \pi}
    \right) < 0
  \end{array} \right.
\end{equation}
Then we can rewrite (\ref{eq:bps}) as a sum over the ``negative'' weights:
\begin{equation}
  I^{\mathrm{BPS}}_{\mathcal{R}} = -\sum_{\{w|w(\vev )<0\}} w
  (\cochar ) = \sum_{i=1}^{\dim(\mathcal{R})}\left[\frac{-1+\sgn(w_i(\vev ))}{2}\right]w_i(\cochar )
\end{equation}
and, using the fact that the weights of any representation sum to zero, we get the final formula:
\begin{equation}
  I^{\mathrm{BPS}}_{\mathcal{R}} = \sum_{i=1}^{\dim(\mathcal{R})}\frac{1}{2}\,\sgn(w_i(\vev ))\, w_i(\cochar )
  \label{eq:bps2}
\end{equation}

Observe that because of the static nature of the solution, the counting of
zero modes on $\mathbb{R}^3 \times S^1$ gives the same result as that of a 3d
theory obtained in the limit of vanishing radius for $S^1$. That might not be
the case for KK monopole backgrounds coming from 4d instanton configurations.
%
%
%
%
%
\tocless\subsubsection{KK monopole background}
\label{app:kkmonopole}
%
%
%
%
%
%
%
When the theory lives on $\mathbb{R}^3 \times S^1$ there is also a second
type of topologically non-trivial background called winding instanton or KK
monopole. This type of solution is the compactification of a standard 4d
instanton on $\mathbb{R}^4$ and can be obtained from a BPS solution by
applying an anti-periodic ``gauge transformation''\footnote{The quotation
marks here are due to the fact that because the transformation is not
periodic, it does not define a proper gauge transformation. In fact, the
transformed solution is not gauge equivalent to the BPS one.} along the $x^4$
direction \cite{Lee:1997vp,Seiberg:1996nz}. In this case the fourth component of the gauge field
cannot be taken to be constant along the compact direction and in fact it defines a non-trivial Wilson
line that wraps the $S^1$. A similar computation to the one in (\ref{app:bpsmonopole}) yields the index of the Dirac operator in a KK monopole background as:
\begin{eqnarray}
  I_{\mathcal{R}}^{\mathrm{KK}} &=& \tr_{\mathcal{R}} \left[ \left\lfloor
  \frac{\vev  L}{2 \pi} \right\rfloor n_0 H_0 \right] + \frac{1}{2} n_0
  \tr_{\mathcal{R}}[H_0 H_0] \nonumber\\
  &=& \sum_{i=1}^{\dim(\mathcal{R})}\frac{1}{2}\sgn(w_i(\vev))w_i(n_0 H_0)
  + 2n_0\frac{T(\mathcal{R})}{\theta\cdot\theta}
\end{eqnarray}
where $H_0$ is the ``affine'' co-root (\ref{eq:affinecoroot}), $T(\mathcal{R})$ is the Dynkin index of the representation and $n_0 \in \mathbb{Z}$ is the KK
monopole charge.
%
%
%
%
%
\tocless\subsubsection{Adjoint representation}
\label{app:adjoint}
%
%
%
%
%
%
%
Here we show that, for a suitable choice of adjoint scalar VEV, a fermion
$\psi$ in the adjoint representation of any gauge group $G$ carries exactly
two zero modes for every unit of BPS or KK monopole charge.

The counting of BPS zero modes is given by the index of (\ref{eq:bps2}) which
can be written as:
\begin{equation}
  I^{\mathrm{BPS}}_{\mathrm{adj}} = \sum_{i=1}^{\dim(G)}\frac{1}{2}\,\sgn(\alpha_i(\vev ))\, \alpha_i(\cochar )
  \label{eq:bpsadj}
\end{equation}
Because we have chosen the VEV of the Higgs field to lie in the fundamental
Weyl chamber (\ref{eq:weylchamber}), the sum ranges over the negative roots with a
minus sign and over the positive roots with a plus sign and the 1/2 in front
takes care of the double counting, therefore we can write:
\begin{equation}
  I^{\mathrm{BPS}}_{\mathrm{adj}} = \sum_{\alpha \in \Delta^+} \alpha
  (\cochar ) = 2 \rho (\cochar )  = \sum_{i = 1}^r 2 n_i
\end{equation}
where $\rho$ is the Weyl vector (\ref{eq:casimir}). The computation of the index for a KK monopole goes as
follows:
\begin{eqnarray}
  I_{\mathrm{adj}}^{\mathrm{KK}} 
  & = & - 2 n_0 \rho \cdot \theta^{\vee} + 2 n_0  \frac{T
  (\mathrm{adj})}{\theta \cdot \theta} \nonumber\\
  & = & 2 n_0 \left( \frac{- 2 \theta \cdot \rho + T (\mathrm{adj})}{\theta
  \cdot \theta} \right) \nonumber\\
  & = & 2 n_0 \left( 1 + \frac{- \theta \cdot (\theta + 2 \rho) + T
  (\mathrm{adj})}{\theta \cdot \theta} \right) 
\end{eqnarray}
where $\theta \cdot (\theta + 2 \rho)$ is the value of the quadratic Casimir
element $C_2 (\mathrm{adj})$ on the adjoint representation. Using the fact that
$C_2 (\mathrm{adj}) = T (\mathrm{adj})$ we finally obtain the desired result:
\begin{equation}
  I_{\mathrm{adj}}^{\mathrm{KK}} = 2 n_0
\end{equation}
%
%
%
%
%
\tocless\subsubsection{$USp$ and $U$ case}
\label{app:uspu}
%
%
%
%
%
%
%
In the main text of this paper we consider $USp (2 N)$ theories with matter in the
fundamental and antisymmetric representations as well as $U (N)$ theories with
matter in the fundamental and adjoint. Here we give the result for the
counting of zero modes for those groups and those representations.\\

In the case of a $USp (2 N)$ gauge group we have one zero mode contribution
coming from the BPS monopole associated to the long simple root:
\begin{equation}
  I_{\mathrm{fund.}} = n_N
\end{equation}
and one for every unit of monopole charge associated to the short simple roots
coming from the antisymmetric representation:
\begin{equation}
  I_{\mathrm{antisymm.}} = \sum_{i = 1}^{N - 1} 2 n_i
\end{equation}
In particular, we observe that in both representations there are no contributions to KK monopole zero modes.\\

In the case of a $SU (N)$ gauge group we have:
\begin{equation}
  I_{\mathrm{fund.}} = n_{i}
\end{equation}
for $i$ the largest integer such that $w_i(\vev ) > 0$ where $w_i$ is the $i$-th weight of the fundamental
representation of $SU (N)$. Therefore there is one contribution coming from the
$i$-th BPS monopole and none coming from the KK monopole.
Similarly, for the case of a $U(N)$ gauge group there are no fundamental zero mode contributions to the KK monopole \cite{Aharony:2013dha}.
%
%
%
%
%
\tocless\subsubsection{Monopole superpotentials}
\label{APPB}
%
%
%
%
%
%
%
Once a 4d duality is reduced on $S^1$ we are in the presence of a new 3d effective duality.
Such a duality has the field and gauge content of the 4d theory and in addition an 
extra superpotential involving the KK monopole.
These effective dualities can then be transformed into more conventional 3d dualities, 
by real mass and higgs flow.
Actually, richer structure of RG flows have been more recently analyzed in \cite{Benvenuti:2016wet},
leading to families of new 3d dualities with non-trivial monopole
superpotentials.
For example it has been shown that the 4d duality of \cite{Intriligator:1995ne}, that relates $USp(2N_c)$ gauge theories with 
fundamentals, can be reduced in this way to a duality between unitary theories with linear monopoles
in the superpotential.
Many of the salient features of these reductions can be captured reducing 
the 4d superconformal index to the 3d partition function.
This reduction gives indeed the 3d identities between the 3d dualities obtained 
from the field theory side.
An important aspect of these reductions regards the constraints between the fugacities
in the superconformal index.
These constraints are necessary in 4d to enforce the constraints imposed by superconformality,
i.e. the vanishing of the beta function or equivalently the anomaly freedom of the R-symmetry current.
The constraints translate in a constraint on the parameters of the 3d partition functions (the real masses
of the associated field theory).
These constraints signal the presence of monopole superpotentials and are usually referred to in the mathematical 
literature as balancing conditions. We encounter such conditions often in our analysis.

Observe that the presence of a monopole in a superpotential fixes the $R$-charge and the abelian flavor charges of 
such a superpotential term, $R[W]=2$ and $F_k[W]=0$ where  the index $k$ runs over the abelian non-R global symmetries.
The charge of a monopole operator with magnetic charge $\cochar $ as in (\ref{eq:defn}) under any global symmetry can be computed
in terms of the charges of the fermions of the theory by a one loop computation.
The quantum correction to the monopole charge is obtained at one loop and it is 
\begin{equation}
\label{monch}
Q_{A}[\mathrm{monopole}] = -\frac{1}{2} \sum_{i} Q_A[\psi_i] |w_i(\cochar )| 
\end{equation}
where $w_i$ is the weight of the $i$-th fermion $\psi_i$ under the gauge group and $Q_A[\psi_i]$ is its charge under the abelian symmetry $A$.\\
%
%
%
%
%
%
%
\subsection{Squashed three sphere partition function}
\label{APPC}
%
%
%
%
%
%
%
Here we provide some general formulas used in the paper.
The partition function 
of a 3d $\mathcal{N}=2$ gauge theory, with gauge group $G$, on a squashed three sphere is given by the general formula
\cite{Kapustin:2009kz,Jafferis:2010un,Hama:2010av,Hama:2011ea}
\begin{equation}
  \label{eq:SquashedSphere}
  Z_{G;k}(\lambda;\vec{\mu}) = \frac{1}{|W|}\int
  \prod_{i=1}^{G} \frac{\mathd\sigma_i}{\sqrt{-\omega_1 \omega_2}}
  \mathe^{\frac{ k \pi \mathi \sigma_i^2}{\omega_1 \omega_2}+\frac{2 \pi \mathi \lambda \sigma_i}{\omega_1 \omega_2}}
  \frac{\prod_I \Gamma_h\left(
  \omega \Delta_I + \rho_I(\sigma)+\widetilde \rho_I(\mu)\right)
  }{\prod_{\alpha \in G_+}\Gamma_h\left(\pm\alpha(\sigma)\right)},
\end{equation}
where the hyperbolic gamma function $\Gamma_h$ (see for example \cite{VdB}) correspond to the contributions of the one loop determinants and are defined as
\begin{equation}
  \label{eq:Gammahvbd}
  \Gamma_h(x;\omega_1,\omega_2) \equiv
  \Gamma_h(x)\equiv 
  \mathe^{
    \frac{\pi\mathi}{2 \omega_1 \omega_2}
    \left((x-\omega)^2 - \frac{\omega_1^2+\omega_2^2}{12}\right)}
  \prod_{j=0}^{\infty} 
  \frac
  {1-\mathe^{\frac{2 \pi \mathi}{\omega_1}(\omega_2-x)} \mathe^{\frac{2 \pi \mathi \omega_2 j}{\omega_1}}}
  {1-\mathe^{-\frac{2 \pi \mathi}{\omega_2} x} \mathe^{-\frac{2 \pi \mathi \omega_1 j}{\omega_2}}}.
\end{equation}
We denote as $b$ the squashing parameter of the ellipsoid defined by the relation 
\begin{equation}
\frac{x_1^2+x_2^2}{b^2} + \frac{x_3^2+x_4^2}{1/b^2}=1
\end{equation}
and define $\omega_1 = \mathi b$, $\omega_2 = \mathi/b$ and $2 \omega \equiv \omega_1+\omega_2$.
In formula (\ref{eq:SquashedSphere}) $\sigma$ and $\mu$  are real quantities, in the Cartan of the gauge and of the flavor symmetry. 
We denoted with an $\alpha$ the positive roots of the gauge group and with $\rho(\sigma) $ and $\tilde \rho (\mu)$ the weights of 
the gauge and of the flavor symmetry respectively, necessary to parameterize the one loop contribution of each chiral field.
The parameter $\lambda$  corresponds to a possible FI term, while the $R$ charge of each chiral field is identified by $\Delta_I$.
The Gaussian factor in the integrand corresponds to the contribution of the classical action and it is identified with the CS term at level $k$.
Possible CS terms involving the flavor symmetries can be turned on and are associated to the contact terms  as discussed in \cite{Closset:2012vg,Closset:2012vp}.

In the paper we mainly studied the partition function of $USp(2N_c)$ gauge theories with $2N_f$ fundamentals and one antisymmetric matter field. By calling $\mu_a$ ($a=1,\dots,2N_f$) the mass parameters
of the fundamentals and $\tau$ the mass parameter of the antisymmetric, the three sphere 
partition function becomes
\begin{equation}
Z_{USp(2N_c)}(\vec \mu;\tau) = 
\frac{\Gamma_h(\tau)^{N_c-1}}{2^{N_c}{ N_c}!(-\omega_1 \omega_2)^{N_c/2}}
\int
\prod_{i=1}^{N_c} \mathd \sigma_i 
\frac{\prod_{a=1}^{2 N_f} \Gamma_h(\tilde{\mu}_a \pm \sigma_i)}{\Gamma_h(\pm 2 \sigma_i)}
\prod_{i<j}
\frac 
{\Gamma_h(\tau\pm \sigma_i\pm \sigma_j)}
{\Gamma_h(\pm \sigma_i\pm \sigma_j)}
\end{equation}
We also used the partition function of $U(N_c)$ gauge theories with 
$N_f$ fundamental flavors and an adjoint.
In this case the partition function has the general form
In this case we define two mass parameters for the flavor, 
$m_a$ and $n_a$ ($i=a,\dots,N_f$), for the fundamentals and the 
anti-fundamentals respectively.
We denote as  $\tau$ the mass parameter of the adjoint. The three sphere 
partition function in this case is
\begin{eqnarray}
Z_{U(N_c)}(\vec m;\vec n;\tau;\Lambda) &=& 
\frac{\Gamma_h(\tau)^{N_c-1}}{ N_c!(-\omega_1 \omega_2)^{N_c/2}}
\int \Big(
\prod_{i=1}^{N_c} \mathd \sigma_i\; \mathe^{\mathi  \pi \Lambda \sigma_i}
\prod_{a=1}^{N_f} \Gamma_h(m_a +\sigma_i) 
\nonumber \\
&\times &
\prod_{a=1}^{N_f} \Gamma_h(n_a -\sigma_i)
\Big)
\times \!\!\!
\prod_{1 \leq i<j \leq N_c}
\frac 
{\Gamma_h(\tau\pm (\sigma_i - \sigma_j))}
{\Gamma_h(\pm (\sigma_i -\sigma_j))}
\end{eqnarray}
where the parameter $\Lambda$ refers to the FI term.
Observe that we did not consider possible constraints among the parameters.
Such constraints have to be added in the presence of non-trivial superpotential interactions,
like monopole superpotentials.

In the analysis we made use of two relevant formulas relating the hyperbolic gamma function.
The first formula 
\begin{equation}
\label{holom}
\Gamma_h(2 \omega - x) \Gamma_h(x)=1
\end{equation}
allows to integrate out pairs of fields associated to superpotential mass terms.
The second formula
\begin{equation}
\label{iout}
\lim_{x \rightarrow \pm \infty} \Gamma_h(x) = \mathe^{-\frac{\pi\mathi}{2} \sgn(x) (x-\omega)^2} 
\end{equation}allows to integrate out fields with a large real mass. The gaussian factor 
in this formula reproduces the CS terms generated on the field theory side.

\bibliographystyle{JHEP}
\bibliography{references}

\end{document}